# High-$T_c$ mechanism through analysis of diverging effective mass for YaBa$_2$Cu$_3$O$_{6+x}$ and pairing symmetry in cuprate superconductors

Hyun-Tak Kim[1]

*Metal-Insulator Transition Lab., Electronics & Telecommunications Research Institute, Daejeon 34129, South Korea*
*School of Advanced Device Technology, University of Science & Technology, Daejeon 34113, South Korea*

**Abstract**

In order to clarify the high-$T_c$ mechanism in inhomogeneous cuprate layer superconductors, we deduce and find the correlation strength not revealed before, contributing to the formation of the Cooper pair and the 2-dimensional density of state, and demonstrate the pairing symmetry in the superconductors still controversial. To the open questions, the fitting and analysis of the diverging effective mass with decreasing doping, extracted from the acquired quantum-oscillation data in underdoped YaBa$_2$Cu$_3$O$_{6+x}$ superconductors, can provide solutions. In this research, the results of the fitting using the extended Brinkman-Rice(BR) picture reveal the nodal constant Fermi energy with the maximum carrier density, a constant Coulomb correlation strength $\kappa_{BR}=U/U_c>0.90$, and a growing Fermi arc from the nodal Fermi point to the isotropic Fermi surface with an increasing $x$. The growing of the Fermi arc indicates that a superconducting gap develops with $x$ from the node (underdoped) to the anti-node (optimally or over-doped). The large $\kappa_{BR}$ results from the $d_{x^2-y^2}$-wave metal-insulator transition for the pseudogap phase in lightly doped superconductors, which can be direct evidence for high-$T_c$ superconductivity. The quantum critical point is regarded as the nodal Fermi point satisfied with the Brinkman-Rice picture. The experimentally-measured mass diverging behavior is an average effect and the true effective mass is constant. As an application of the nodal constant carrier density, to find a superconducting node gap, the ARPES data and tunneling data are analyzed. The superconducting node gap is a precursor of $s$-wave symmetry in underdoped cuprates. Furthermore, the half-flux quantum, induced by the circulation of $d$-wave supercurrent and observed by the phase sensitive Josephson-π junction experiments, is not shown due to 'anisotropic or asymmetric effect' appearing in superconductors with trapped flux. The absence of $d$-wave superconducting pairing symmetry is also revealed.

## 1. Introduction

Since the discovery of the high-$T_c$ cuprate layer superconductor in 1986, numerous studies have been conducted and theories also developed on the subject [1-8]. In particular, the parent materials of cuprate high-$T_c$ superconductors are known to be antiferromagnetic Mott insulators, which are formed by a strong on-site critical Coulomb interaction between carriers in metal. The Mott insulators have an electronic metallic structure with half filling and semiconducting properties. When a low concentration of either hole or electron charges is doped into Mott insulators, the pseudogap phase appears instead of the antiferromagnetic insulator. Moreover, it has been proposed that the pseudogap phase arises from charge- and/or spin-density waves occurring in layered organic, iron pnictide, transition metal dichalcogenide, 2H-NbSe$_2$ [9-12]. At higher doping, a metal phase emerges over $T_c$ and superconductivity appears below $T_c$ [1-8]. The electronic structures in the superconducting state have also been elucidated by angled resolved photoemission spectroscopy experiments[13,14] in k (or energy)-space, and by the tunnel effect in real-space [13-15].

Nevertheless, the mechanism underlying the high-$T_c$ superconductivity has not yet been clarified, because underdoped cuprate superconductors are highly inhomogeneous due to severe non-stoichiometry arising from a strong oxygen disorder induced by doping inhomogeneity [16,17], although the optimal or overdoped crystals are more homogeneous [18,19], have a large continuous cylindrical Fermi surface [20], and show $s$-wave symmetry [21-23]. The homogeneous region has been investigated at about 3 nm in an underdoped Bi$_2$Sr$_2$CaCu$_2$O$_{8+x}$ (BSCCO or Bi-2212) superconductor [24] and at approximately 100 nm in a stoichiometric, overdoped YBa$_2$Cu$_3$O$_7$ (YBCO) superconductor [18,19], utilizing scanning tunneling microscopy. It is, therefore, very difficult to measure the homogeneous region as a means of ascertaining the true value. Inhomogeneity hinders the analysis of the intrinsic physical value, because the obtained experimental value is averaged and depends on the measurement region; for instance, the carrier density increases as hole doping increases from the underdoped region to the optimal region although the carrier density in metal is constant due to the metal condition of one electron per atom. The true nature of the superconductor should be deduced through the detailed analyses of the observed and averaged data. Based on these deductions, the mechanism of high-$T_c$ superconductivity, therefore, becomes identifiable. Thus, an understanding of inhomogeneity in underdoped cuprate superconductors, rather than theoretical studies, is vitally important for revealing the mechanism of high-$T_c$ superconductors.

In order to understand the mechanism of high-$T_c$ superconductors based on inhomogeneity, several important unexplored or controversial issues must be resolved, including the relationship between the pseudogap and superconductivity, the identity of the pairing symmetry (or order parameter) of the superconducting gap, and the magnitude of the density of states as an effective mass connected to the correlation between carriers. Above all, the cause of the nodal constant Fermi point (or velocity) must be understood, along with the interrelationship between the issues [25-28]. Analysis of the diverging effective mass (DEM) measured in the metal state of a superconductor is a very good example on the metal-insulator transition related to correlation theories for explaining the mechanism of high-$T_c$ superconductivity [29-33].

---

[1] htkim@etri.re.kr; htkim580711@gmail.com

A several years ago, an important experimental study was published on underdoped YaBa$_2$Cu$_3$O$_{6+x}$ (YBCO) crystals, suggesting the cause of the nodal constant Fermi point and the pairing-symmetry identity [20,32,33]. A diverging effective mass (DEM) in Fig. 5 was obtained from quantum-oscillation measurements in underdoped YBCO crystals [20,32,33], and the results showed an increasing effective mass with decreasing $x$, although the extent of the metal region also decreased (Fig. 1) [32,33]. This observation is anomalous, because the oscillation and the effective mass coming from the metal phase should decrease (see first paragraph of section 3-2). Thus, the DEM has been unfitted and its physical meaning remains unclear,[32,33] although the presence of a small pocket was suggested at the node [34-36]. Furthermore, since the DEM data were measured at the node, the clarification of the physical meaning of this data can disclose important information on the mechanism behind high-$T_c$ superconductors and the pairing symmetry for cuprate superconductors.

Here, this paper is composed of three sections and their subsections in the main text except for introduction. In the first (2-x) and second (3-x) sections, we fit and analyze the DEM data (Fig. 3C in reference 32 and Fig. 5 in this paper) extracted from the quantum oscillation measured at the node in underdoped YBCO$_{6+x}$ crystals [32], using the extended Brinkman-Rice picture to explain the MIT (metal-insulator transition) in inhomogeneous systems. We then use the fitting results and subsequent analysis to shed light on the physical meaning of the DEM. In the third section (4-x), the nodal constant Fermi energy and the identity of the pairing symmetry are revealed through further analysis of the La$_{2-x}$Sr$_x$CuO$_4$, YBCO and Bi$_2$Sr$_2$CaCu$_2$O$_{8+x}$ (BSCCO or Bi-2212) ARPES data. Additionally, in subsections of 4-3, we reanalyze the well-known observations of the half-flux quantum that were measured through phase sensitive experiments of YBCO, using a Josephson-$\pi$ junction, which has been suggested as strong evidence of $d_{x^2-y^2}$-wave symmetry of the Cooper pairs for the cuprate superconductors.

## 2. An MIT theory and Its Application

### 2-1. The extended Brinkman-Rice(BR) picture (EBR)

In Figs. 1a and b, we briefly introduce the extended BR picture, explaining the MIT as a two-phase model applied to an inhomogeneous system of a local metal region and an insulator region, both contained within the measurement region (black circle) [30,31,37]. This model has two conditions: (1) a fractional effective charge $e'=\rho e$ with band filling $0<\rho=n_{metal}/n_{total}\leq 1$, defined by an average of metal carriers, $n_{metal}$, to the measurement region, where $n_{total}$ is the total number of lattices in the measurement region, and (2) a fractional effective Coulomb energy of $0<U/U_c=\kappa_{BR}\rho^2<1$, where $U=\rho^2 U'$ is defined (Fig. 1c), $U'=\kappa_{BR}U_c$ is the on-site repulsive Coulomb energy, $U_c$ is the critical Coulomb energy, and $\kappa_{BR}=U/U_c$ at $\rho=1$ is the true correlation strength of the Brinkman-Rice picture [29]. Due to the fact that metal has an electronic structure of one electron per atom (lattice) in real space and half filling in k space ($\rho=1$) (Fig. 1a), the physical meaning of $U$ is that the metal region widens as $\rho$ increases (Fig. 1b).

Thus, the diverging effective mass was calculated as

$$\frac{m^*}{m} = \frac{1}{1-\left(\frac{U}{U_c}\right)^2} = \frac{1}{1-\kappa^2\rho^4} \quad \text{-------} \quad (1)$$

[30,31]. The detailed derivation is given in a reference [31]. Eq. (1) is plotted in Fig. 2, which shows two divergences at $\rho=1$ (BR-DEM) and $\kappa_{BR}=1$ (Kim-DEM). The BR-DEM is the true effect, while Kim-DEM is the measurement (average) effect and represents the percolation phenomenon. BR-DEM and Kim-DEM are terms that were first used to distinguish

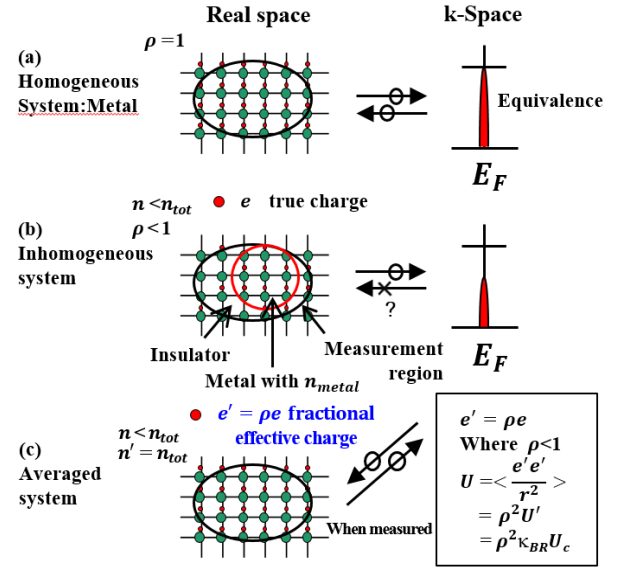

**Figure 1. An idea for the extension of the Brinkman-Rice (BR) picture.** (a) Metal has the electronic structure of one electron (small red dot) per atom (green large dot) in real space and is homogeneous in the measurement region. It has a constant Fermi energy (half filling) in k-space. A relationship of equivalence between real space and k space is satisfied. (b) In an inhomogeneous system, the metal region is part of the measurement region. The equivalence relationship is not satisfied. (c) The number of carriers in the local metal region in Figure (b) is divided by the number of lattices (atoms) in the measurement region. This indicates that the carrier charges are averaged for all of the lattices. The averaged fractional effective charge is given as $e'\equiv(n_{metal}/n_{tot})e=\rho e$ and $0<\rho\leq 1$. The system is consistent with Figure a and equivalence. In this case, the fractional effective on-site repulsive Coulomb energy is defined as $U=\rho^2 U'=\rho^2\kappa_{BR}U_c$, where $\kappa_{BR}$ is the correlation strength in the Brinkman-Rice (BR) picture.

two divergences in a previous paper [38]. Two divergences deduce the metal-insulator transitions between the Mott insulator at set $\rho$=1 and $\kappa_{BR}$=1 and the divergence points. One is named the Brinkman-Rice MIT 1 (BR-transition 1) between max($\kappa_{BR}$)<1 and $\kappa_{BR}$=1 at $\rho$=1, and the other is a Brinkman-Rice MIT 2 (BR transition 2) between max($\rho$)<1 and $\rho$=1 at $\kappa_{BR}$=1. The BR-transition 2, at the hole impurity concentration, $\Delta\rho$, between the maximum $\rho_{max}$ near $\rho$=1 and $\rho$=1 (the yellow region in Fig. 2b) when $\kappa_{BR}$=1, is called the hole-driven MIT (or impurity-driven MIT). This MIT (BR-transition 2) occurs by excitation between a doped insulator denoted by red dot and metal in Fig. 2b; red dot line. The doped insulator is formed when a critical hole concentration (red dot in Fig. 2b) of $\Delta\rho_{critical}=\Delta n/n_{tot}$, 1 x $10^{18}$ cm$^{-3}$, corresponding to an impurity of $\Delta\rho_{critical}$=0.018% [37], as determined in a representative strongly correlated material, $VO_2$, is doped into a Mott insulator (black dot in Fig. 2b, 2c); the critical concentration of $\Delta\rho$ differs from Mott's criterion of the minimum carrier density obtained from the minimum conductivity in main band [39]. The insulator is defined as an impurity-doped Mott insulator with an impurity band undergoing the MIT (or insulator-to-metal transition (IMT)) at a high temperature [Fig. 4c in reference 40] (Fig. 2d), which is different from the known antiferromagnetic Mott insulator without impurity band (Fig. 2c) and with both an electronic structure of one electron per atom in real space and the Hubbard bands in k-space (Fig. 2c). In more detail, when temperature, pressure, chemical doping, or a mechanical stress-induced strain is applied to the impurity-doped Mott insulator, the impurity charges in the impurity band of the insulator are excited, whereupon the main Hubbard bands formed by the critical Coulomb interaction between carriers (Mott insulator) or the Mott gap disappears (breakdowns) and transitions to metal, which is the impurity-driven MIT [40,41]. The MIT point is regarded as an MIT instability and corresponds to the quantum critical point, since this model is defined at $T$=0 K. The process of the IMT can be accompanied with the phase separation induced by a local inhomogeneous doping specifying over (or critical) or less doping of the critical impurity concentration $\Delta\rho_{critical}$ driving the MIT; the less $\Delta\rho_{critical}$ doping generates a semiconductive phase regarded as the pseudogap one (the yellow region in Fig. 2b). Furthermore, the impurity-induced MIT(BR-transition 2) can explain the universal MIT behavior suggested in an experimental MIT research of 'universality of pseudogap and emergent state in lightly doped Mott insulators' [42]. Note that Mott considered the impurity-driven MIT and an indirect transition of band crossing, but the former was not completed [39,43]. Note that an enhancement in mobility for a nanowire field-effect transistor was explained by, $\mu^*= (e\tau/m)(1-\rho^4)$, using the BR and the EBR pictures [44].

Furthermore, as for the percolative phenomenon, the continuous increase of the metal region induced by MIT, percolation occurs in accompaniment with the impurity-driven-indirect Mott-MIT(the BR-transition 2). Two types of percolations based on $\rho$ are introduced. One type occurs when the extent of the metal region increases in the constant measurement region (Case 1 in Fig. 3), while in the other, the measurement region enlarges with respect to the constant metal region (Case 2 in Fig. 3).

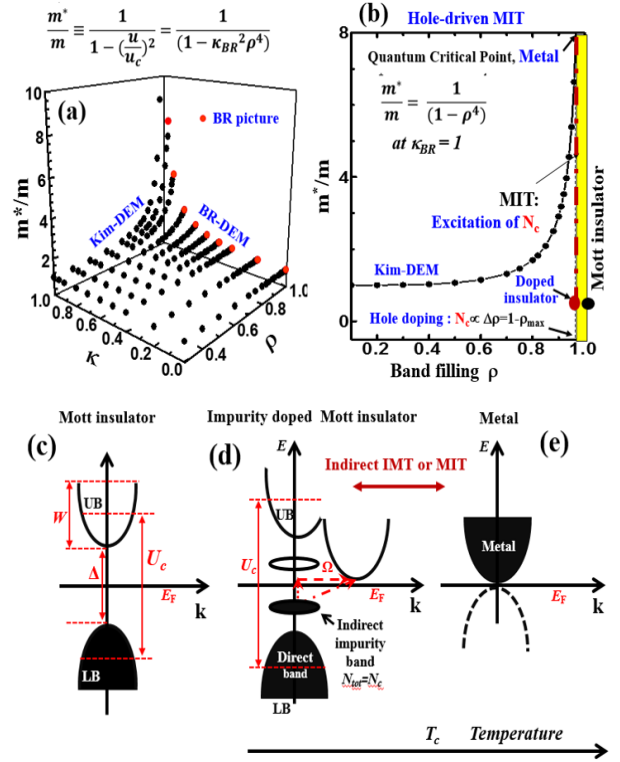

**Figure 2.** Drawing of Eq. (1) in the extended BR picture [30,31]. Eq. (1) shows two divergences of the BR diverging-effective mass (BR-DEM, red-bullets) at $\rho$=1 and the Kim-DEM at $\kappa_{BR}$=1. The BR-DEM and Kim-DEM terminology was first used in a previous paper [38]. (b) shows the hole-driven MIT (BR-transition 2) at $\kappa_{BR}$=1. The yellow portion represents the semiconductor regime and $\Delta\rho$ at red dot corresponds to hole doping and the impurity doped insulator. The metal-insulator transition (MIT) is shown at red dot line between doped insulator and metal specifying excitation. $n_c \equiv N_c \propto \Delta\rho$ is the doping concentration for the MIT. The quantum critical point is given at the transition point. (c) The antiferromagnetic Mott insulator with the Mott gap of $U_c$ is assumed at $U/U_c$=1 in the BR picture (Fig. 2c), as denoted by black dot in Fig. 2b. Fig. 2d shows a band structure of an impurity-doped Mott insulator (red dot in Fig. 2b) with both the main Hubbard bands for direct transition and an impurity (or extrinsic semiconductor) band for indirect transition. UB – upper Hubbard band, LB – lower Hubbard band [40], $E_F$ – Fermi level, $\Delta_{direct}$ – energy gap for direct transition, $\Delta_{act}$ – activation energy for indirect transition, $\Omega$ – thermal phonon. Impurity concentration $N_{tot}=N_c$ is proportional to $\Delta\rho=N_{tot}/n_{tot}$ in the EBR picture, where $n_{tot}$ is the carrier density in the main Hubbard band. The IMT (or MIT) is indirect between Fig. 2d and Fig. 2e, which names the indirect Mott MIT [40]. For a strongly correlated insulator of $VO_2$, the IMT criterion, $N_{tot}=N_c$, is 0.018% [37].

## 2-2. Application of the MIT theory : The MITs in $YBCO_{6+x}$

In order to explain the MIT in YBCO, the hole-driven MIT (BR-transition 2) is applied to YBCO. A structure of $YBa_2Cu_3O_6$ without oxygen chain is shown (Fig. 4a). As oxygen into $YBa_2Cu_3O_6$ is doped, a CuO chain is formed as $YBa_2Cu_3O_{6+x}$ (Fig. 4b) [45]. The CuO chain is an antiferromagnetic insulator and an impurity-doped semiconductor with a direct Mott gap of 2 eV [46] and an indirect gap (impurity level) of 0.12-0.16 eV [47]. The $CuO_2$ plane in the $YBa_2Cu_3O_{6+x}$ below $x$=0.4 is still an insulator with an energy gap of 1.5~4 eV [46]. $YBCO_6$ is an intrinsic Mott insulator with the tetragonal structure [48] (red point in Fig. 3c) and $YBCO_{6+x}$ is defined as a doped Mott insulator, as shown at red point in Fig. 4c. When the indirect Mott MIT (BR-transition 2) induced by oxygen doping (excitation of impurity charges in the impurity level) occurs in the CuO chain near $x \approx 0.45$ (Figs. 2 and 4c), $YBCO_{6+x}$ experiences the SPT from tetragonal to orthorhombic structure, as shown in Fig. 4c, and displays suddenly superconductivity at the orthorhombic structure [48-50]; this MIT is similar to that in $VO_2$ [37,40,41]. Then, the insulating $CuO_2$ plane in the tetragonal $YBCO_{6+x}$ becomes metal (in the orthorhombic $YBCO_{6+x}$) inducing superconductivity. Moreover, when an oxygen concentration of over $x \approx 0.45$ is doped, a still Mott insulating $YBCO_{6+x}$ phase (CuO chain) with doping $x$ not reaching a critical doping, $x \approx 0.45$, due to doping inhomogeneity continuously changes into a metallic (or superconducting at low temperature) phase, which is accompanied with the SPT from tetragonal to orthorhombic. It follows from the continuous MIT and SPT with doping that this is the percolation shown in reference [32]. Note that the Mott MIT in CuO chain near $x \approx 0.45$ can occur before (or without) the SPT, which remains to be experimentally proved.

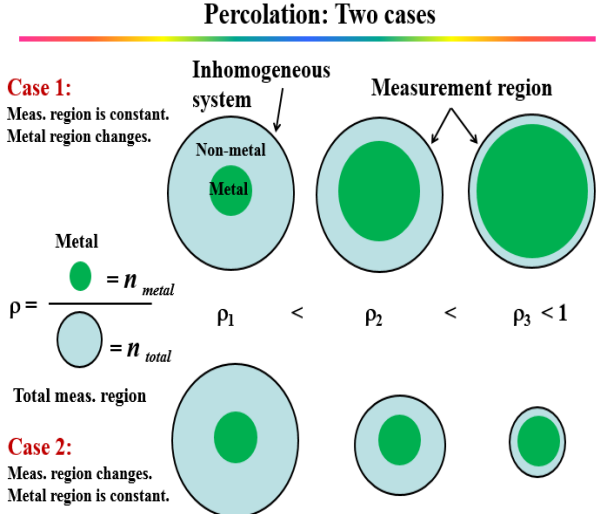

**Figure 3. Percolation models.** Case 1 shows a percolation in which the metal region increases in the constant measurement region. In Case 2, the measurement region changes to a constant metal region. However, $\rho = n_{metal}/n_{tot}$ is the same in both cases, where $n_{metal}$ is the number of free carriers in metal region and $n_{tot}$ is the number of lattices (atoms) in the measurement region.

## 3. Fitting of the DEM extracted from quantum-oscillation data in $YBCO_{6+x}$

### 3-1. Understanding quantum oscillation

Quantum oscillations were measured at the node in underdoped $YBCO_{6+x}$ crystals with $x$ in strong magnetic fields at 1.5 K [32]. These strong magnetic fields broke the superconducting state and triggered a normal state. The oscillation signal at the node was much larger than that observed at the anti-node and had a comparatively weak dependence on the magnetic field [32].

Before the quantum oscillation data can be fitted, the experimental method must be understood. Therefore, the oscillation frequency was extracted from the conductivity via the Fourier transformation [32], and the conductivity was measured with a contactless method in a pulse magnet, using a tunnel diode oscillator circuit with resonance frequencies. This method indicates that the quantum oscillation is a physical quantity and is dependent on the magnitude of $k$-space (carrier density), such as the Fermi energy.

Moreover, it was suggested that the Brinkman-Rice (BR) picture [29] can fit the measured DEM [32] (Fig. 5), but that the BR picture alone cannot fit the data measured in inhomogeneous YBCO crystals, because the picture is applied to a homogeneous system with an electronic structure of one electron per atom. Thus, the extended-BR picture in Eq. (1) is a necessary alternative [30,31].

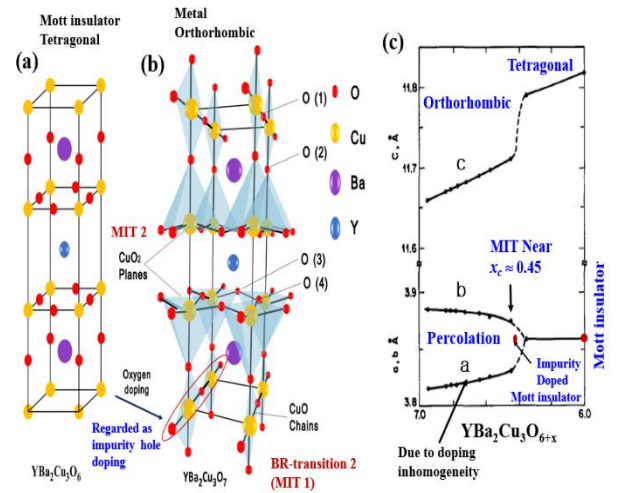

**Figure 4. Structures of YBCO and MIT in YBCO.** (a) Atomic structure of $YBa_2Cu_3O_6$. (b) Atomic structure of $YBa_2Cu_3O_{6+x}$ doped by oxygen. The structure has a CuO chain with a Mott gap of 2 eV [46] and an indirect gap of 0.12 -0.16 eV [47]. The CuO undergoes the BR-transition 2 (MIT 1) at $N_c$ (Fig. 2b). The metallic $CuO_2$ plane (MIT 2) appears after the structural phase transition (SPT) from tetragonal to orthorhombic. The SPT occurs after the MIT 1 in the CuO chain. (c) Refined crystallographic cell parameters for $YBa_2Cu_3O_{6+x}$ prepared by Zr gettering at 440°C. This was cited from Fig. 5 in Cava *et al*. [48].

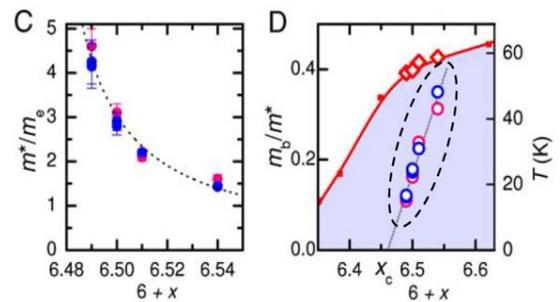

**Figure 5. Diverging effective mass (left) and linear behavior in dot line of inverse effective mass (right).** These are in Figs. 3C and 3D of Reference, Sebastian *et al*. [32], which is the figures to be explained here.

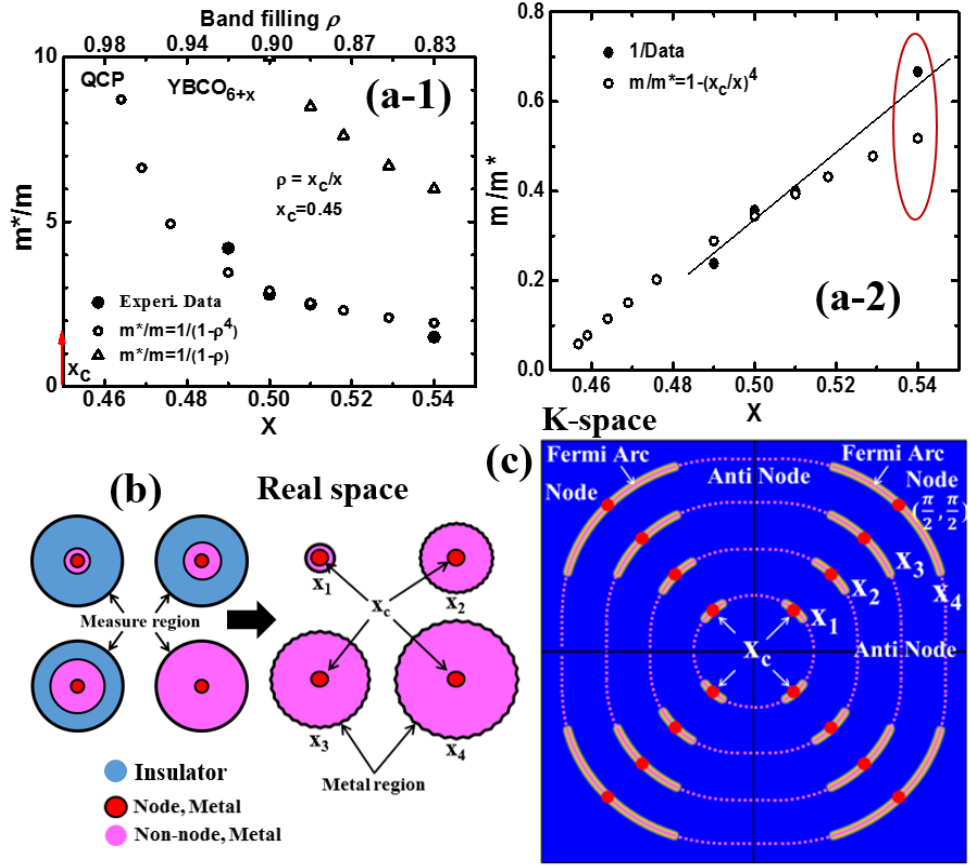

**Figure 6. The fitting of the diverging effective mass.** (a-1) The diverging effective mass data (black dot) extracted from quantum oscillation data in Fig. 3C by Sebastian, *et al.*, (Fig. 5) [32] are closely fitted by Eq. (1) when $\kappa_{BR}$=1. Upper *x* axis is expressed as a function of band filling $\rho$. QCP means the quantum critical point. (a-2) The inverse effective masses, $m/m^*$, of Fig. (a-1) are shown. At *x*=0.54 far from the transition point, the fitting of Eq. (1) was deviated from the experimental data showing the linear behavior, which requires more detail measurements. (b) A schematic percolation diagram showing an increase of metal region (pink) with *x* (left figure corresponding to case 1 in Fig. 3). Based on $\rho_i = x_c/x_i$, as *x* (pink) increases, such as $x_1 \approx 0.49 < x_2 \approx 0.50 < x_3 \approx 0.51 < x_4 \approx 0.54$ [32], $\rho$=(red/pink) decreases (right figure corresponding to case 2 in Fig. 3); $\rho_1 > \rho_2 > \rho_3 > \rho_4$ with respect to $x_1 < x_2 < x_3 < x_4$ is a percolation. The extent of the metal region (red dot) at the node is constant because the numerator of $\rho$ is $x_c$. (c) A schematic figure of the Fermi point and surface in k-space based on an analysis of Figs. 6a and 6b. Because the extent of the metal region (pink region) increases as *x* increases (Fig. 6b), in k-space, the Fermi surface grows from the Fermi point at the node (red dot) to the Fermi arc (pink arc) at the anti-node. However, the magnitude of the Fermi point at the node (red dot) is constant irrespective of increasing of *x*. There is no Fermi energy (or dot or surface) at the anti-node (0, π).

### 3-2. Key idea for fitting of the DEM

First of all, we analyze a phenomenological linear behavior with inverse $m^{-1} \propto x$, as shown in Fig. 3D of reference 32 and Fig. 5. YBCO$_6$ was known as the Mott insulator with half filling. YBCO$_{6+x}$ of a notation used in reference 32 is a hole-doped crystal with a deviation $\delta$ from half filling. Thus, $\delta$=1-*x* is defined. A theoretical effective mass given as a function of $\delta$, $m^{-1} \propto \delta = x_c$-*x*, where '1' implying half filling is replaced by $x_c$ (an explanation is given below) was approximately calculated in a condition of $U \gg U_c$ [51]; this formula is different from $m^{-1} \propto x$ expressing an experimental behavior in Fig.5. The theoretical formula indicates the increase of the effective mass with increasing *x*, which corresponds to the percolation Case 1 in figure 3. This is opposite to the experimental result, as shown in Figs. 3C and 3D of reference 32 and Fig. 5. Moreover, in previous researches, for correlated materials such as Sr$_{1-x}$La$_x$TiO$_3$, the inverse $m^{-1} \propto \delta \equiv 1$-*x* deviated from the experimental data near the transition point [30]. Thus, the DEM and the linear behavior in Fig. 5 are not explained by the previous theoretical model with DEM in the effective mass and linear behavior in the inverse effective mass.

In order to fit the DEM data in Fig. 5 [32], a new concept is necessary. Alternative is to use Eq. (1) in the EBR picture because of the deviation ($\delta \neq 0$) from half filling; the BR picture can be applied in a condition of half filling (the deviation $\delta$=0). For the fitting, the following assumption is required: inhomogeneous YBCO crystals have two phases as a metal and an insulator (pseudogap phase). The metal phase is also composed of two types of metals at the node ($\pi/2, \pi/2$) and at the non-node including anti-node (0, π) (Fig. 6b); this is supported by the well-known ARPES experiments on growing Fermi surface according to doping. Since the quantum oscillation is defined in *k*-space, the insulator phase without the Fermi surface does not emerge in *k*-space. The metal phase appears on the Fermi surface in the Fourier transformation, and two metal phases should therefore be considered, one at the node and the other at the non-node (anti-node). Because the measurements[32,33] were taken at the node, the carrier density of the metal phase at the node denotes $n_{metal\ node}$ in $\rho$, and $n_{total}$ includes the carrier densities at both the node and the anti-node (all regions apart from the node). At the node with $\rho=1$, $n_{metal\ node}$ amounts to half filling. Thus, $n_{tot}=n_{metal\ node}+n_{metal\ non-node}$ is given. When the extended BR picture and this YBCO system are compared, $n_{metal\ non-node}$ corresponds to an insulator component in the extended BR picture (Fig. 1b) because $n_{metal\ non-node}$ does not logically contribute to the quantum-oscillation data measured at the node. In other words, because $\rho=n_{metal}/n_{tot}$ should be defined, the extent of the metal region at the node, $n_{metal\ node}$, corresponds to the critical doping, $x_c \equiv x_{c\ node}$, and the size of the measurement region, $n_{total}$, which includes the metal regions at both the node and non-node including anti-node, is regarded as $x$ in YBCO$_{6+x}$, because $x = x_{c\ node} + x_{non-node}$. The reason that $n_{metal\ node}$ corresponds to $x_c$ is because the MIT first occurs at node, when the minimum doping concentration, $x_c$, inducing the MIT is doped into the doped Mott insulator. Thus, $\rho \equiv n_{metal}/n_{tot} \approx x_c/x$ is given. As $x$ increases, $\rho$ decreases, and vice versa (Fig. 6b). This is the percolation phenomenon corresponding to Case 2 in Fig. 3. The above assumption is a key idea in fitting.

### 3-3. Fitting of the DEM with the EBR picture and the nodal Fermi point

We fit the experimental DEM data in Fig. 5 with both the fitting idea and Eq. (1) with conditions of $\kappa_{BR}=1$ and $\rho \approx x_c/x$. The fitting result is shown in Figs. 6(a-1) and 6(a-2). Four data points are closely fitted to Eq. (1) except for the fitting at 0.54 far from the transition point (see red circle in Fig. 6(a-2) showing the inverse effective mass); whether the deviation from the linear behavior at $x=0.54$ is true or not needs to be studied because experimental data have experimental errors from a standard value. The fitting at deviating region may not be important because the region where data are saturated has no the physical meaning. Therefore, we suggest that Eq. (1) fits the experimental DEM data.

This fitting reveals that the extent of the metal region (numerator, $x_c$) at the node is constant regardless of $x$ (right figure in Fig. 6b; this follows the percolation model of case 2 in Fig. 3), although the size of the metal region (denominator, $x$) increases with increasing hole-doping $x$ (left figure in Fig. 6b; this follows the percolation model of case 1 in Fig. 3), which implies an increase in the Fermi energy, indicating that the Fermi energy at the node is constant. Moreover, an MIT at the critical hole-doping $x_c$ occurs at the node. Thus, $x_c$ corresponds to $N_c=\Delta\rho$ in Fig. 2b and $n_{metal\ node}$ is formed. Based on the fit, it would be expected that, as hole-doping $x$ increases, the Fermi surface would grow from the nodal Fermi point to the anti-nodal Fermi arc (Fig. 6c). The measured Fermi arc [52-56] is an intermediate process in reaching the isotropic Fermi surface and the origin of the nodal Fermi point is that the MIT begins to occur at the node, rather than at the anti-node [25]. The nodal Fermi point (or velocity) has previously been demonstrated [25-28], as has the constant maximum carrier density at the nodal Fermi point (or velocity) [25-28]. Furthermore, the inverse effective mass, $m/m^* = 1-(x_c/x)$, does not fit the experimental data due to the inverse $x$, as shown in Fig. 6(a-1). Note that the universal nodal Fermi velocity [27] was suggested on the basis of a constant slope below 70 meV in energy-momentum space, but, because the energy range is far from the Fermi surface, this suggestion was criticized [26]. Vishik et. al reasserted a doping-dependent nodal Fermi velocity [26]; this is different from our suggestion mentioned here.

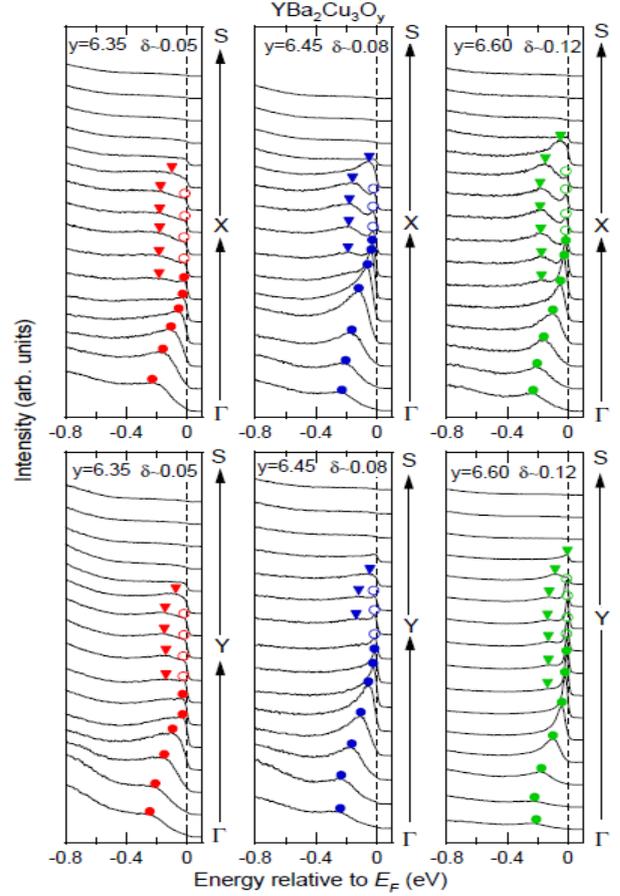

**Figure 7. The pseudogap at the anti-node.** ARPES spectra measured at 10 K for hole-doped non-superconducting single crystals (y=6.35, 6.45, 6.60) of YBCOy. Pseudogaps of approximately -0.2 eV are clearly shown at the X and Y antinodes (inverse bold triangles ▼, ▼, ▼) (also see Fig. 9). The value is independent of hole doping. This was cited from Figure 4.3 in Yagi's PhD paper [61].

The nodal Fermi points are formed with the maximum number of carriers (red dots in Fig. 6c, the metal part of the constant size in Case 2 of Fig. 3) caused by the $d$-wave MIT (Figs. 7-9). In the nodal points, the inhomogeneity disappears and the homogeneous metal phase is formed at the node where $\rho=1$ [57]. The nodal point is regarded as the quantum critical point, which is defined as a point where the MIT occurs at $T=0$ K, which is satisfied with the BR picture specifying the true constant effective mass. The diverging behavior of the measured effective mass in Fig. 6a is an average (measurement) effect, which is fitted by Eq. (1). This nodal Fermi point is very important for the formation of superconductivity. Moreover, when the pairing symmetry is explored, the carriers at the nodal Fermi point should be investigated, whether or not they are bound. This investigation is given through the analysis of experimental data in following sections.

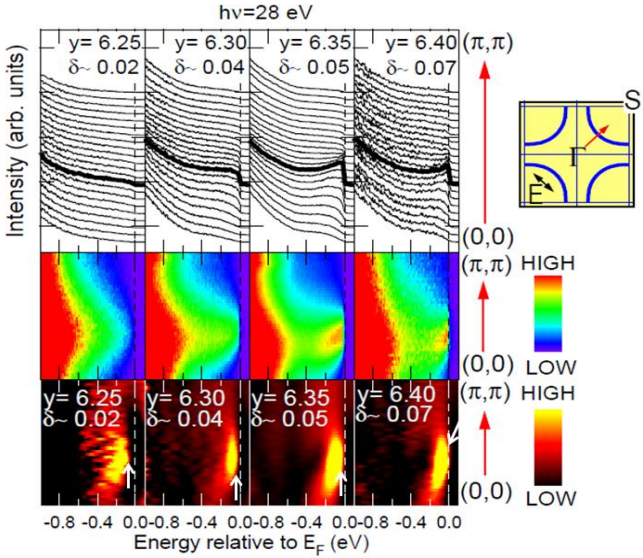

**Figure 8. The *d*-wave MIT from the pseudogap to the metal at the node.** ARPES spectra measured at 10 K at the node for lightly doped single crystals of YBCOy with y=6.25 (pseudogap of -0.1 eV), 6.30, 6.35, and 6.40, without $T_c$. This data was cited from Figure 5.1 in Yagi's PhD paper [61].

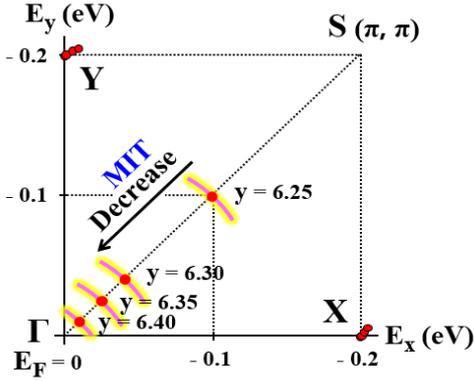

**Figure 9. The doping dependence of pseudogaps in Figs. 7 and 8.** As y increases, the pseudogap at the anti-node remains unchanged (Fig. 7), while the edge of the pseudogap (white arrows, bottom Figures in Fig. 8) decreases at the node. The MIT occurs at the node (*d*-wave MIT). This shows that the pseudogap has the *d*-wave symmetry.

### 3-4. Constant correlation strength $\kappa_{BR}$ and experimental confirmation

In order to obtain a value of $\kappa_{BR}=U/U_c$, we assume that $m^*/m=1/(1-\rho^4)$ (Kim-DEM) at $\kappa_{BR}=1$ is approximately equal to $m^*/m=1/(1-\kappa_{BR}^2)$ (BR-DEM) at $\rho=1$ near the diverging point of $\rho=1$ and $\kappa_{BR}=1$ (Fig. 2a, Fig. 6a-1). Thus, $m^*/m \approx 4.5$ at $x=0.49$ is given[32,33] and $\kappa_{BR}$ was determined to be approximately 0.89. However, because $x$ can approach $x_c$, the effective mass can greatly increase. Therefore, we suggest a constant of $\kappa_{BR}>0.9$, which indicates that the true effective mass is always constant, irrespective of doping concentration, as long as the metal region is measured with a tool on a measurement region that is smaller than the metal region in Fig. 1b [31]. The constant effective mass was theoretically predicted [30] and observed across the entire underdoped regime of the phase diagram for $La_{2-x}Sr_xCuO_4$ and $YBa_2Cu_3O_y$ [58].

A large value of $\kappa_{BR}>0.9$ can be evidence that the metal phase at the node is strongly correlated. This strong correlation is attributed to the *d*-wave MIT of the pseudogap phase [57], which can be the cause of the high-$T_c$ in the cuprate systems. The pseudogap shows *d*-wave characteristics [59] and competes with the superconductivity [60].

Moreover, we speculate two roles in respect to the effects of the strong correlation $\kappa_{BR}=U/U_c>0.9$ contributing to the high-$T_c$ superconductor. One is an attractive potential $V'=(-V_o + U) < 0$, induced by an interaction to bind carriers in metal, which becomes much smaller when $U$ becomes significantly larger (easy tunneling), allowing for the tunneling of superconducting carriers, such as Cooper pairs, between the attractive potential. The other is the increase of the 2-dimensional density of states when the effective mass greatly increases as a function of $U$ (the BR picture).

The fitting result is verified with experimental data obtained from lightly hole-doped $YBCO_y$ insulators (semiconductor or pseudogap). Figures 7 and 8 show the angle-resolved-photoemission-spectroscopy (ARPES) spectra measured at 10 K for single $YBCO_y$ crystals [61]. A summary of these two figures is provided in Fig. 9. As the doping y increases, the pseudogap at the anti-node does not change (Fig. 7), while the pseudogap rapidly decreases and finally disappears at the node (Fig. 8). This is the *d*-wave MIT at the node, indicating that the pseudogap has $d_{x^2-y^2}$-wave symmetry. This is also the origin of the Fermi arc. Therefore, the Fermi arc as fitting result in Fig.6c is proven by the ARPES data.

### 4. Application to high-$T_c$ superconductors

#### 4-1. Analysis of superconducting node gap

Many researchers believe that the pairing symmetry of cuprate superconductors is $d_{x^2-y^2}$-wave, based on the observations of both the zero-bias-conductance peak (ZBCP), which implies the absence of the superconducting gap at the node in ARPES and tunneling experiments, and the half-flux quantum, which is measured at the Josephson-$\pi$-junction point. The ZBCP is regarded as a characteristic of $d_{x^2-y^2}$-wave symmetry. Note that the $d_{x^2-y^2}$-wave gap is defined by $\Delta_d \equiv \Delta_s Re(e^{i2\phi}) = \Delta_s \cos 2\phi$, where $\Delta_s$ is the isotropic *s*-wave gap and $\phi$ is the azimuthal angle. However, many measurements of the superconducting gap have shown an absence of the ZBCP in tunneling experiments [15,62-64]. This discrepancy is obviously problematic. It must be reexamined whether the nodal Fermi points (red dots in Fig. 6c) with maximum (or constant) carriers ($\rho=1$), as explained by section 3-3, contribute to the formation of the superconducting gap or whether they remain unchanged below $T_c$. If the nodal Fermi points remain in the metallic state, the absence of the node gap should be observed as the ZBCP and the pairing symmetry would be *d*-wave, otherwise, it would be *s*-wave with a nodal superconducting gap. Therefore, we investigate the presence of the superconducting node gap, in order to resolve this problem on the basis of tunneling and ARPES data.

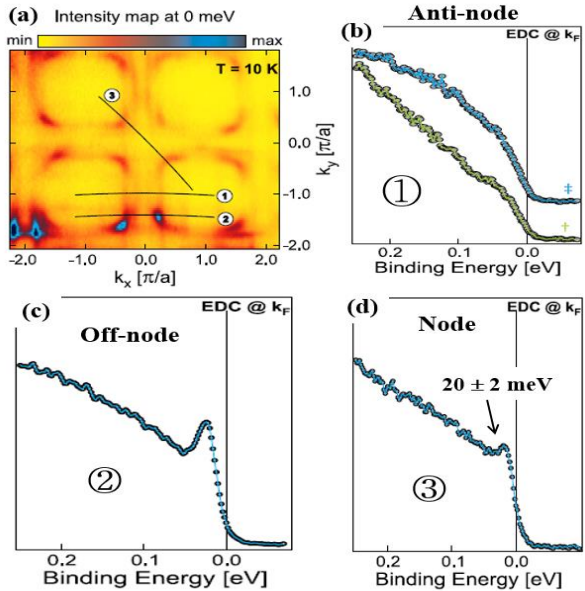

**Figure 10. The superconducting node gap.** (a) Near-Fermi level $E_F$, spectral intensity map acquired at a photon energy of 70 eV with energy integration of ARPES spectra over ± 5 meV window about $E_F$ for a YBCO film deposited by PLD. (b, c, d) ARPES spectra acquired at cut (1), (2), and (3) in k-space as indicated by the lines in (a). The energy distribution curves (EDC) at $k_F$ and T = 10 K are shown for the cuts indicated in the spectra (dashed blue and green lines). The small energy gap is shown at the node. This data was cited from the Sassa's PhD paper [65].

### 4-1-1. Analysis of ARPES data

The important ZBCP is not observed in ARPES spectra, because the spectra are given by a product of the imaginary part of the one-particle Green function and the Fermi distribution function. However, the ARPES spectra can reveal the existence of a node gap because a s-wave superconductor has the superconducting gap at the node. In order to find the presence of a small gap at the node, we analyze well-measured ARPES data. The ARPES data measured in an underdoped YBCO$_{7-\delta}$ crystal showed the absence of a gap at the anti-node, a large gap at the off-node, and a small superconducting gap at the node (Fig. 10) [65], suggesting the presence of the Fermi arc in the normal state. In addition, a superconducting small node gap of $20 \pm 2$ meV was also observed in a YBCO$_{7-\delta}$ crystal, although the $d_{x^2-y^2}$-wave gap at anti-node, which is interpreted as the pseudogap shown in Figs. 7-9, was seen [66]. The ARPES data, which was measured in underdoped and optimally-doped single La$_{2-x}$Sr$_x$CuO$_4$ (LSCO) crystals, and proven by the Fermi arc (Fig. 11) [55,56], show small superconducting gaps at the node (Figs. 11d-e and Fig. 12a). Near optimal doping, the gap anisotropy with small and large gaps is also exhibited in ARPES (Fig. 11f). The magnitude of the node gap decreases with an increasing doping concentration (Fig. 11 (d-f) and Fig. 12a), which was also observed in lightly-doped cuprates [67]. The presence of the node gap has been considered with various scenarios, including the possible effects of chemical disorder, electronic inhomogeneity, and a competing phase [67]. It has been suggested that the node gap is generic in all cuprate superconductors [62]. The node-gap shrinkage with $x$ has also been interpreted as a measurement (average) effect of inhomogeneity (see Fig. 11g-i) [25]. Moreover, a small gap near the node was shown to disappear near $T_c$ [55,56], suggesting that the node gap is both superconducting and intrinsic [25,28]. The gap anisotropy was also observed in cuprate superconductors [13,54,68] and, when measured in an underdoped BSCCO, showed that the pseudogap has $d$-wave symmetry (Fig. 12b). In addition, the ARPES data measured with a high-resolution system for an underdoped BSCCO crystal also showed a small gap at the node [54], as shown in Fig. 13. These data indicated that the hump at the anti-node disappears at the node. It is therefore deduced that the node gap is formed during the change from the homogeneous metal state (with maximum carriers in nodal Fermi points) to the gap state at low temperatures (Figs. 11 and 12). The magnitude of the small gap at the node was measured as approximately $\Delta_{sup} \leq 10$ meV at the node for YBCO, BSCCO [13,54,55], Ca$_{2-x}$Na$_x$CuO$_2$Cl$_2$, Nd$_{2-x}$Ce$_x$CuO$_4$ [67], and LSCO with x=0.15 (Fig. 11f).

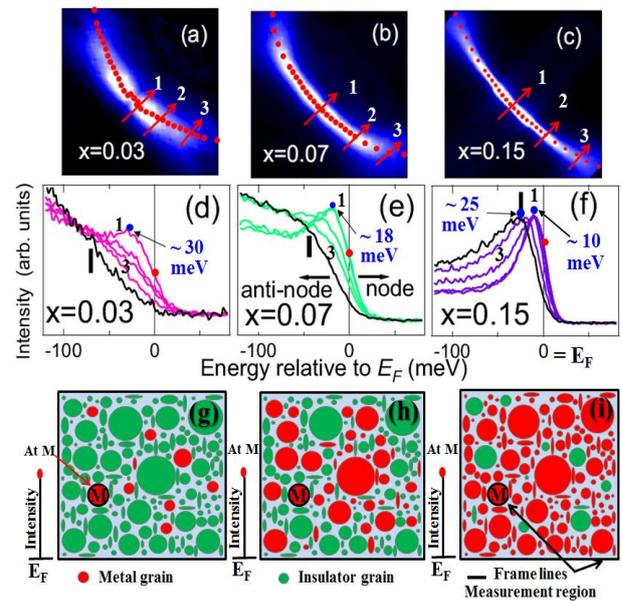

**Figure 11. The ARPES spectra showing the Fermi arc** were measured at 20 K for single crystals of La$_{2-x}$Sr$_x$CuO$_4$ (LSCO) [55,56]. $T_c$ was 14 K for the crystal with x=0.07 and 39 K with x=0.15 [55,56]. The crystal with x=0.03 had no $T_c$. Figs. a-c show the Fermi arcs. Figs. 11d-f depict the energy-distribution curves (EDC) measured near each arrow (Figs. 11a-c). The black vertical bars (Figs. 11d-f) correspond to anti-nodal EDC, and the vertical bars represent the energy position of the anti-nodal gap [55]. As hole doping increases, the Fermi energy also increases (red dots at $E_F$ in Fig. 11d-f and the energy gap decreases at the node (as designated by 1 in Fig. 11d-f). These phenomena correspond to percolation Case 1 (Fig. 3 and Figs. 11g-i). In particular, the larger superconducting gap presented in the crystal with both x=0.03 and no $T_c$ is shown, as indicated by 1 and a bold black dot (Fig. 11d). This suggests the local presence of microscopic superconducting grains (Fig. 11g), even though $T_c$ is not measured macroscopically. If the measurement region is reduced to grain size, the gap size and $T_c$ will be equal for all samples (as symbolized by M in Figs. 11g-i) [25]. Figs. 11(a-f) were cited from Yoshida, *et al.* [56].

### 4-1-2. Analysis of the tunneling energy gap

The tunneling energy gaps measured with $dI/dV$ for cuprate superconductors, including YBCO in real space, have shown U and V shapes without the ZBCP on the a-b plane [63,64,69] and a V+ZBCP shape [69,70]. The U shape corresponds to

arXiv:1710.07754 [cond-mat.supr-con] 21 Oct. 2017

the energy gap of the *s*-wave superconductor, but V and V+ZBCP shapes are associated with the energy gaps of *d*-wave superconductors. The V-shape gap has a gap structure without carriers at the Fermi energy (zero-bias voltage) [69,70], suggesting the presence of a small nodal gap, since the tunneling gap is an average of energy gaps between the node and the anti-node in k (or energy)-space. If there are carriers as evidence of *d*-wave at the node, the ZBCP should appear at the Fermi energy. Moreover, the hidden pseudogap was smaller than the superconducting gap without the ZBCP and was observed in the stoichiometric (overdoped) $Sr_{0.9}La_{0.1}CuO_2$ [19]. The pseudogap screened the superconducting gap near zero bias voltage, but the absence of the ZBCP suggests the presence of a node gap. Thus, the small node gap and the tunneling gap without the ZBCP are not explained by *d*-wave, but rather by *s*-wave, which is to say, the node carriers (nodal Fermi point shown in Fig. 6c) contribute to the formation of the superconducting gap.

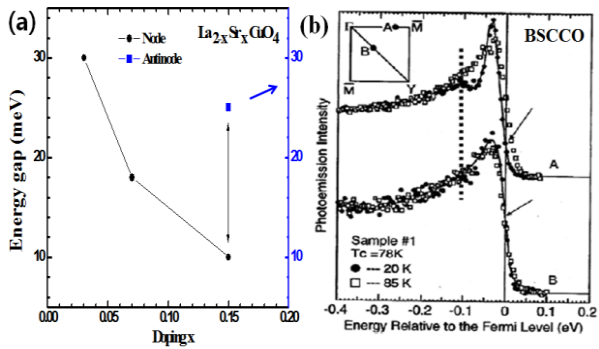

**Figure 12. Superconducting energy gaps in Fig. 11 and the gap anisotropy.** (a) As doping and $T_c$ increase at the node, the energy gap decreases. In the case of an anti-node, the anti-node gap exists only at $x=0.15$. The gap anisotropy is shown in the crystal with $x=0.15$ (Fig. 11f). The energy gap at the node is approximately 10 meV (Fig. 11f). Moreover, the superconducting energy gap was displayed for the crystal of $x=0.03$ without $T_c$. The sample did not exhibit superconducting gaps at the anti-node (Fig. 11d-e), which is evidence of the absence of the Fermi energy at the anti-node. This is consistent with the fitting result (Fig. 6c). (b) The gap anisotropy measured in an underdoped BSCCO by ARPES. The hump, regarded as the pseudogap at the antinode (A curve), disappears at the node (B curve); this is illustrated by the dotted line, which indicates the *d*-wave insulator (pseudogap)-to-metal transition [25,28]. This reveals that the pseudogap is *d*-wave. The Fermi-level position at curve A rises at curve B. This result indicates that the superconducting gap (large gap) at the antinode is smaller at the node; this is the gap anisotropy. This was cited from Shen *et al.* [68].

### 4-1-3. Coupling constant between the superconducting gap and $T_c$

Furthermore, the strength of the coupling constant between the superconducting gap and $T_c$ is deduced. At the optimal doping, the coupling constants are revealed as $b \equiv 2\Delta_{sup}/k_B T_{c,max} \approx 4.0$ at $\Delta_{sup} \approx 10$ meV and $T_{c,max} \approx 93$ K for YBCO and $b \approx 5.95$ at $\Delta_{sup} \approx 10$ meV and $T_c \approx 39$ K for LSCO data (Fig. 11f). It is known that for low-$T_c$ superconductors, $b \approx 3.5$, but the values are slightly over $b \approx 4.0$ for high-$T_c$ superconductors. The difference in $b$ values may arise from the correlation strength, $\kappa_{BR}$, increasing the density of states. Namely, the high-$T_c$ cuprate layer superconductors have larger $\kappa_{BR}$ values than the low-$T_c$ superconductors, because of the *d*-wave electronic structure of the pseudogap phase in the a-b plane layer of $CuO_2$ of the cuprate superconductors. The mechanism underlying the formation of the superconducting node-gap is summarized in Fig. 14.

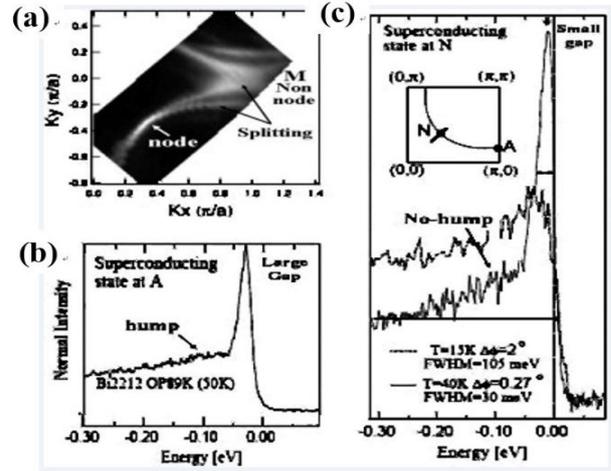

**Figure 13. Node gap observed by ARPES**. (a) Image showing node and anti-mode (or non-node) measured by ARPES for Pb-doped BSCCO. At the node, the intensity is very bright, while at the anti-node, it is low and separated. This was observed by Bogdanov, *et al.* [53]. (b) For the BSCCO ARPES data, the large gap at anti-node A is regarded as the pseudogap. (c) The small gap at node N, corresponding to the superconducting gap and its magnitude, is approximately 10 meV. Figs. (b) and (c) were cited from Kaminski, *et al.* [54]. Fig. 13 was cited in a reference [28].

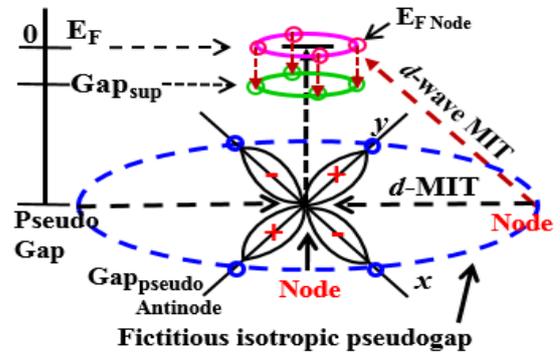

**Figure 14. A mechanism for the formation of the node gap.** The $d_{x^2-y^2}$ electronic structure can be formed when the metal-insulator transition (MIT) occurs at the node in an isotropic pseudogap structure (blue dashed ring). The small pink circles in the large pink circle are regarded as the nodal Fermi points made by the *d*-wave MIT near doping $x_c$. The pink circle is the Fermi surface formed by increased doping. The red-dashed arrow indicates that bound charges in the pseudogap potential at the node are excited to the Fermi energy, due to the *d*-wave MIT (conceptual indication). The small green circles in the large green circle are superconducting intrinsic gaps at the node when the nodal Fermi points become a superconductor (pink circle -> green circle) [25,28]. The green ring is the isotropic superconducting *s*-wave-like gap resulting from the Fermi arc at optimal doping. If the superconducting energy gap has $d_{x^2-y^2}$-wave-pairing symmetry, the *d*-wave-MIT should occur at the anti-node. However, this research does not support the *d*-wave pairing symmetry. The constant maximum carrier density at the nodal Fermi point (or velocity) was first disclosed [25].

### 4-1-4. Analysis of *d*-wave gap measured by ARPES

We analyze the typical $d_{x^2-y^2}$-wave-superconducting gap, as suggested and measured for an underdoped BSCCO crystal (Fig. 15) [14]. Figure 15a shows the typical $d_{x^2-y^2}$-wave gap with the ZBCP, implying zero gap at 45°, which was measured with the laser ARPES [14]. Below 29°, the energy gap curve follows the *s*-wave form at the anti-node, but shows the *d*-wave form above 31°. At 102 K > $T_c$≈92 K (Fig. 15b), the gaps still exist below 24° and the ZBCPs are distinct over 24°. The existence of the gap over $T_c$ deviates from the characteristic disappearance of the superconducting gap at $T_c$. Therefore, the anti-node gap is interpreted, not as a superconducting gap but as a pseudogap. The existence of the superconducting gap near the anti-node is anomalous on the basis that the Fermi arc is formed near the node in an underdoped crystal. A small superconducting gap is expected to exist at the node, but the trace of the node gap is not shown. The gap shape is V type (Fig. 15d). When the gaps at each angle in Fig. 15a are also integrated, the integrated gap is the same as the typical tunneling gap seen with the ZBCP (Fig. 15d) [69,70]. The ZBCP is known to emerge at sites of impurity [71], at an interface of the break junction [72], in a-axis or (110) oriented films [73,74], at a grain boundary [75,76], or at the vortex [77] in underdoped samples with a pseudogap. This data indicates that the ZBCP is caused by extrinsic effects. When it is taken into account that the superconductor is generated in the clean metal phase without extrinsic effects, the pseudogap phase of the ZBCP and the clean metal phase inducing the superconductor may be separated; thus, the two phases do not interact with each other. It is considered that this pseudogap to be different from the pseudogap contributing to the formation of the node gap mentioned in sections 3-4. The latter may be on the same plane of $CuO_2$ forming the node gap. This reveals that the anti-node gap and the typical *d*-wave-gap behavior are interpreted as characteristics of a pseudogap with *d*-wave symmetry, as seen in Figs. 7 and 9. It therefore is concluded that the pseudogap phase, not the superconducting phase, is overwhelmingly dominant in the measurement region; there is very little superconducting phase in the measurement region. This may lead to the absence of a node gap. This conclusion is different from that of the authors [14]. Moreover, from the above analysis, it follows that the *d*-wave gap, which was measured in the BSCCO that exhibits the same behavior as Figs. 15(a, b, c) and has been suggested as strong evidence of *d*-wave symmetry [78], came from the pseudogap phase. Additionally, recent ARPES results have shown that the pairing symmetry deviates from a simple *d*-wave [13,55].

### 4-2. *S*-wave components observed by tunneling experiments

**Fully gapped superconductivity** has been reported in a nanometer-size $YBa_2Cu_3O_{7-\delta}$ island enhanced by a magnetic field [79]. This research also revealed that the energy required to add an extra electron depends on the parity (odd/even) of the excess electrons on the island and increases with magnetic field. This is not explained in terms of the $d_{x^2-y^2}$-wave scenario with unpaired carrier pockets at nodes without a gain of energy. The presence of an *s-wave*-like isotropic superconducting full gap was deduced by fitting the orthodox transport theory from I-V curve data that was measured by a single electron transistor with both an underdoped superconducting (103) YBCO quantum dot and a normal conductor/insulator/superconductor/insulator/normal-conductor tunnel structure. The gap was shown to be three orders of magnitude smaller than the maximum value of the *d*-wave gap. This implies the presence of the *s*-wave like superconducting gap in the underdoped YBCO quantum dot crystal. We suggest that the *s*-wave gap corresponds to the node gap of the crystal, as has been revealed in previous sections of our research.

**For optimally- or over-doped $Bi_2Sr_2CaCu_2O_{8+\delta}$ bi-single crystals,** the ratio $J_c^J/J_c^S$ of the c-axis twist junction critical current density to that across either single crystal part was measured to be unity, independent of the twist angle and the junction area ratio, $A^J/A^s$ [22]. Observing a c-axis twist Josephson junction, $J_c^J/J_c^S$ vs. the c-axis twist angle was closely fitted to *s*-wave, rather than *p* or *d* waves [22]. The authors deduced that the order parameter had an *s*-wave component with a non-vanishing Fermi-surface average near $T_c$. Thus, the optimal- or over-doped crystal has the *s*-wave like Fermi surface developed from the node, as described in a previous section of our research. We note that this is a representative study suggesting strong evidence of *s*-wave symmetry soon after the discovery of the high-$T_c$ superconductors.

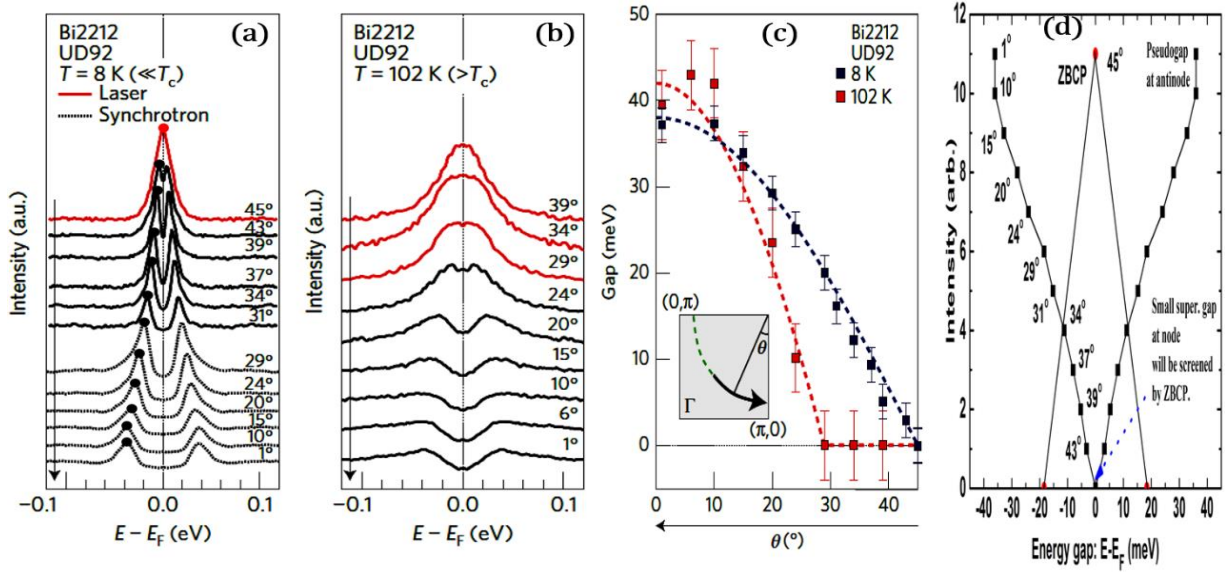

**Figure 15. Analysis of the *d*-wave superconducting gap.** (a) and (b), The angle and temperature dependence of the anti-node gap for an underdoped Bi-2212 crystal, as measured by ARPES. The ZBCP at 45° in Fig. a was measured by laser ARPES. In Fig. b, the gap shapes below 20° are shown over $T_c$, which indicates that the antinodegap is the pseudogap. (c) The arranged angle dependence of the data is shown in Figs. a and b. The angle dependence of the antinode gap reveals that the energy gap follows typical *d*-wave symmetry. Figs. (a-c) were cited from Hashimoto, *et al.* [14]. (d) Summary of the gap curves in Fig. a. The gap points correspond to the bold dots at the gap peaks in Fig. a. The gap shape in Fig. d is V type which is the typical behavior of the pseudogap. When the constant carriers, as revealed in Fig. 6, are present and contribute to the superconducting small gap at the node, the small gap should exist at the node.

**In regards to the c-axis Josephson research on $Bi_2Sr_2CaCu_2O_{8+x}$ (BSCCO, Bi-2212),** BSCCO has no CuO chains but shows a distortion of the $CuO_2$ layers due to the BiO layers [80]. Nevertheless, the distortion of the layers, diagonal to the Cu-O bonds, should lead to a diagonally distorted *d*-wave order parameter. The investigation of the order parameter through c-axis tunneling experiments is very important and the experimental results revealed that the c-axis Josephson tunnel junctions between underdoped BSCCO single crystals and Pb showed a Fraunhofer diffraction pattern (FDP) with a Gaussian curve near zero magnetic field for 3 samples, indicating that the critical current density was homogeneous over the whole junction area. Shapiro steps in external microwave fields were observed at voltages of $nf\Phi_o$ (*f* is a microwave frequency and *n* is the natural integer), showing the Josephson current. These revealed the presence of an *s*-wave component in the crystals. This research does not show evidence of *d*-wave symmetry, being absent in both the Josephson current at zero field and the Shapiro step at voltages of $nf\Phi_o/2$. These were the characteristics measured in the optimally- or overdoped crystals. The authors noted that the used crystals were strongly underdoped, although the BSCCO crystals were annealed in oxygen. Moreover, analyses of both the $I_cR_n$ products and $I_c$ vs. temperature suggested that the ratio of the *s*-wave component to the *d*-wave component was approximately $10^{-3}$ [80]. However, in the background of the suggestion, the *d*-wave superconducting gap was extracted from a fitting of a mixed theory of *s+d* or *s+id* symmetries not a direct measurement. The magnitude of the extracted *d*-wave gap was justified by being compared to tunneling gaps in literatures. Because the tunneling gap measured in a strongly underdoped crystal takes an average of anti-node gap (pseudogap revealed here or no superconducting gap, as shown in Fig. 10b) and node gap (superconducting gap) in k-space, the magnitude of the obtained *d*-wave superconducting gap is not correct. Thus, the ratio of $10^{-3}$ is invalid. The authors indicated that the *s*-wave superconducting gap obtained from $I_cR_n=\pi\Delta_{sup}/2e$ was very small [80]; an average value of the measured $I_cR_n$ was 2.8 μV. This is evidence supporting the presence of the superconducting node gap (small gap), as suggested in previous sections, because the strongly underdoped crystals have no superconducting gap at anti-node. Moreover, similarly, a small superconducting node gap of approximately 4 meV was measured by using the ARPES [26].

**In observations of the Josephson pair c-axis tunneling between $YBCO_{7-\delta}$ and Pb [23]**, the Josephson tunneling currents were observed on low leakage Pb/insulator/YBCO and Pb/insulator/$Y_{1-x}Pr_xBCO$ tunnel planar junctions with c-axis tunneling. The very clean Fraunhofer diffraction pattern with a Gaussian curve near zero magnetic field was measured, which is suggested as direct evidence of *s*-wave symmetry. The used YBCO crystals were annealed for a period of ~1 week in oxygen, suggesting that they were overdoped. The *s*-wave-like Fraunhofer pattern is a characteristic of overdoped samples with a well-developed Fermi surface, as suggested in our research.

**For optimally-doped $Nd_{1.85}Ce_{0.15}CuO_{4-y}$ (NCCO) and underdoped $La_{1.89}Sr_{0.11}CuO_4$ (LSCO) single crystals,** the ZBCP was investigated by point-contact tunneling spectroscopy [81]. The (100) and (110) NCCO crystals did not show the ZBCP, while the (110) LSCO crystal did. The (110) plane of the crystals corresponds to node one. The cuprate compounds readily react with atmospheric $H_2O$ and $CO_2$ to form insulating hydroxides and carbonates on the surface at grain boundaries [82]. The hydroxides and carbonates make cuprate characteristic degradation and growth an unexpected scattering factor. Therefore, the authors used nonaqueous chemical etching to maintain a good sample surface. Nevertheless, the surface degradation of the LSCO crystals from exposure to air was greater than that of the NCCO crystals, due to the existence of reactive alkaline earth elements in LSCO [82]. For this reason, it is clear that the observed ZBCP is merely a surface effect. In particular, the obtained tunneling data in the optimal NCCO crystals showed the best fit to the anisotropic $s$-wave model. The authors suggested the presence of the $s$-wave component in the optimally-doped NCCO, in contrast to the existence of the $d$-wave component for the observation of the ZBCP in (110) LSCO (x=0.11). We mention that the severe surface degradation in the LSCO crystals, screening the intrinsic crystal characteristics, had an influence on the occurrence of the ZBCP.

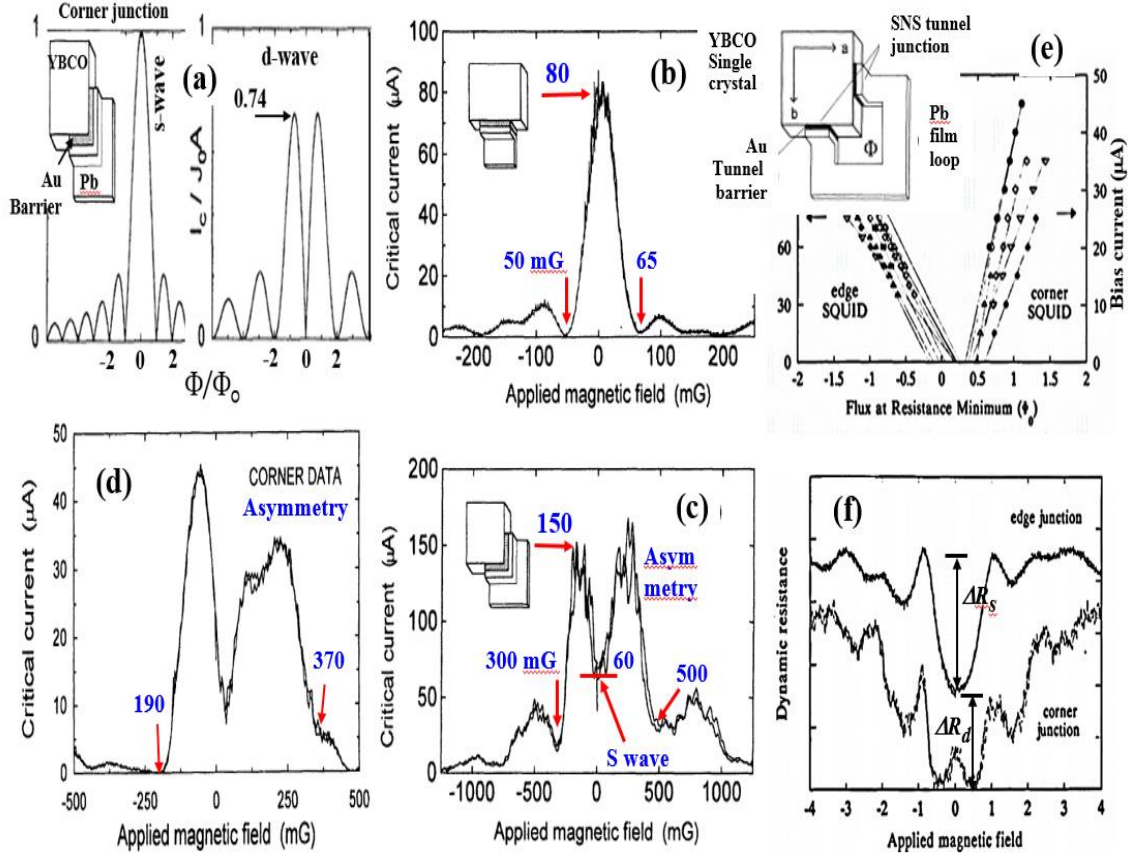

**Figure 16. Trap of magnetic flux and asymmetry.** (a) The theoretical Fraunhofer diffraction pattern (FDP) of $I=I_o|\sin(\pi\Phi/\Phi_o)/(\pi\Phi/\Phi_o)|$ for $s$-wave and $I=I_o|\sin^2(\pi\Phi/2\Phi_o)/(\pi\Phi/2\Phi_o)|$ for $d$-wave. (b) The FDP measured at the YBCO-Au-Pb Josephson edge junction. (c) The FDP measured at the YBCO-Au-Pb Josephson corner junction. (d) The FDP measured at the YBCO-Au-Pb Josephson corner junction when a trapped flux exists. Figs. a-d are cited from Wollman *et al.* [87]. (e) Flux vs. current measured at the YBCO-Au-Pb edge and corner SQUIDs. The current at ring structure of the corner junction is given as $I=I_o|\cos((\pi\Phi/\Phi_o)+phase/2)|$ where phase is zero for $s$-wave and $\pi$ for $d$-wave. (f) Dynamic resistance vs. applied magnetic field measured for the YBCO-Au-Pb edge and corner SQUIDs. Figs. e and f were cited from Wollman *et al.* [88].

Recently, in superconducting copper-oxide ($CuO_2$) monolayer films on $Bi_2Sr_2CaCu_2O_{8+\delta}$ (Bi-2212) known as the $d$-wave superconductor [83,84], the U-shaped STS conductance spectra ($dI/dV$), regarded as evidence of $s$-wave symmetry, were measured on the $CuO_2$ plane, while V-shaped spectra on the BiO surfaces, revealed as the pseudogap state, were also observed. The authors [83,84] asserted this measurement is anomalous, although the possible explanations on this measurement were given [84]. We unveil our analysis on the anomalous measurement. The copper-oxide monolayer films were prepared on the BiO surfaces of the cleaved Bi-2212 crystals by a state-of-the art ozone molecular beam epitaxy technique. When oxide crystals are grown in ozone, the crystals are born as an oxygen rich overdoped sample, because ozone translating into activated oxygen reacts easily with oxides. Moreover, the experimental evidence that these films are overdoped shows a low $T_c$ of approximately 91 K lower than that seen in an optimally-doped sample (usually $T_c$>120 K). The authors proposed that the U-shaped spectra originated from the hole-doping to the adjacent BiO/SrO reservoir layers for the formation of the two-dimensional hole liquid confined in the $CuO_2$ layers. We suggest that the $CuO_2$ monolayer was an overdoped sample with the full Fermi surface showing a U-shaped gap below $T_c$ because the sample was grown in ozone.

Zhang [84] proposed three scenarios to explain the formation of the U-shaped gap on the $CuO_2$ without doubting the $d$-wave symmetry of the cuprate. The first scenario is that the charge carriers on the monolayer $CuO_2$ are heavily

**12 / 19**

transferred into the inner layers, so that the electronic structure of the monolayer is essentially the same as that in the sandwiched $CuO_2$ planes in the bulk Bi-2212. The second is that the charge carriers on the $CuO_2$ monolayer are far away from the copper oxide planes in the bulk Bi-2212, and that the pairing is intrinsic. Within this scenario, the U-shaped gap would then be a new superconducting state. The third is that the monolayer is a good metal, and that the observed superconductivity is due to the proximity effect of the pairing from the substrate cuprate. The scenarios are different from the *s*-wave model suggested here and are based on extrinsic effect. Thus, we propose the investigation of the Fermi surface of the $CuO_2$ monolayer.

Recently, the possibility of the anomalous U-shaped measurement was reported through a theoretical analysis using a two-band model [85]. It was suggested that the U shape can be measured when the hole concentration in the monolayer $CuO_2$ plane is very large [85]. We indicate that this suggestion is consistent with our analysis because the overdoped crystals have the completed Fermi surface of the ring-type as well as a lot of carriers.

**The coherent terahertz (THz) emission technique** is a very good tool for the exploration of the superconducting node gap for an underdoped superconductor with the mixed gaps of both a pseudogap with a large gap and the superconducting node gap, as suggested in this study, because the emitted THz frequency depends on the magnitude of the superconducting gap ($eV=\Delta_{sup}=Nhf/2$, $N$ is the number of intrinsic Josephson junctions and $f$ is the THz frequency) [86]. For a resonator with the thermally–managed disk stand-alone mesa sandwich structure ($N=2$) of an underdoped BSCCO crystal, a maximum radiation frequency of $f=2.4$ THz was observed [86], which translates into the superconducting gap of $\Delta_{sup}\approx 9.9$ *meV*. $\Delta_{sup}$ agrees with the magnitude of the superconducting node gap measured for a BSCCO single crystal in ARPES (Fig. 13c) [54]. Thus, the observed gap is regarded as an intrinsic superconducting gap with *s*-wave symmetry.

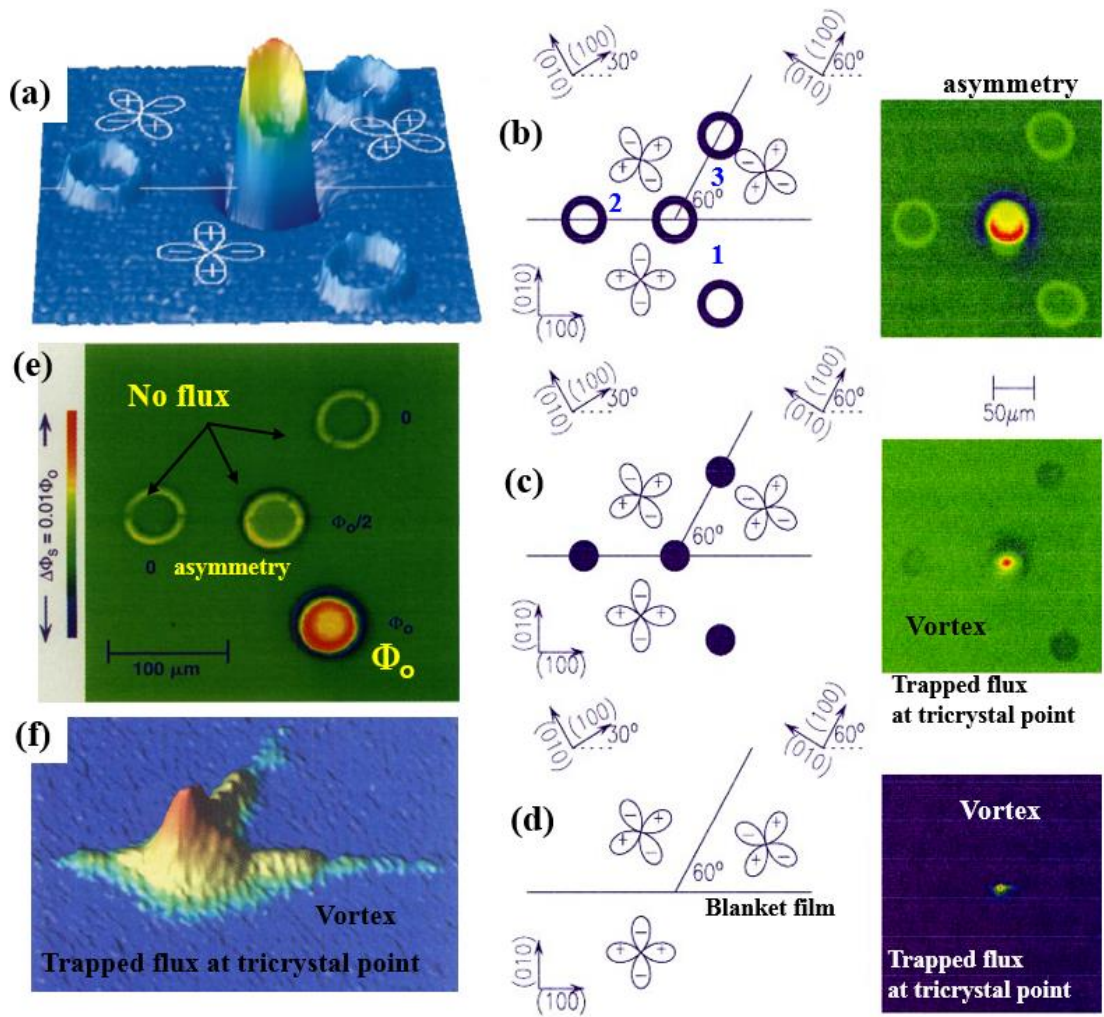

**Figure 17. Images of trapped magnetic flux.** (a) and (b) The half magnetic flux imaged by a SQUID microscope at an underdoped YBCO superconducting ring on the tricrystal substrate. (c) and (d) flux images for YBCO films covering the tricrystal point regarded as the trapped flux induced by the vortex coming from a tricrystal point structure. In Fig. (d), a blanket YBCO film covers the substrate, including the tricrystal point. Figs. a-d are cited from Tsuei *et al*. [89,90]. (e) Flux image of ring type measured in an optimal-doped single-layer tetragonal $Tl_2Ba_2CuO_{6+\delta}$ superconducting ring without a junction [96]. (f) Flux image of large intensity at the tricrystal point for the blanket $Tl_2Ba_2CuO_{6+\delta}$ film, which is regarded as trapped flux in a vortex induced by the tricrystal structure [96]. Figs. e and f were cited from Tsuei *et al*. [96].

## 4-3. Analysis of phase sensitive experiments observing the half-flux quantum

As strong evidence of *d*-wave symmetry in the superconducting gap of cuprate superconductors, the half-flux quantum (spontaneous magnetization) was proposed and measured at the corner junction using a twinned YBCO-Pb SQUID [87] and YBCO-Au-Pb Josephson junctions [88], and in the tricrystal YBCO superconducting ring using Josephson-$\pi$-junction tunneling [89,90], and at the YBCO-Pb SQUID [91-94]. Because the observation of the half flux is a key to resolving the superconducting mechanism in cuprate superconductors, the experiments need to be shed new light on.

### 4-3-1. YBCO-Au-Pb Josephson corner junction

Above all, we reanalyze the experimental data showing the half-flux quantum obtained from the magnetic field modulation of the YBCO-Au-Pb Josephson corner junction [88]. Figure 16a shows the theoretical Fraunhofer diffraction pattern (FDP) of $I=I_o|\sin(\pi\Phi/\Phi_o)/(\pi\Phi/\Phi_o)|$ for *s*-wave and $I=I_o|\sin^2(\pi\Phi/2\Phi_o)/(\pi\Phi/2\Phi_o)|$ for *d*-wave. The *s*-wave form has a large diffraction pattern at zero magnetic flux, while the *d*-wave form includes the divided symmetry pattern at positive and negative fluxes. The magnitude of the *d*-wave FDP is about 74% of that of the *s*-wave. The critical current measured at the edge junction displayed the *s*-wave FDP and its magnitude was about 80 μA (Fig. 16b). The pattern width is shown to be approximately 115 mG (50 + 65). However, in the corner junction (Fig. 16c), the critical current FDP exhibited *d*-wave characteristics, but had a finite value at zero field regarded as characteristic of *s*-wave and a width of 300 mG at the negative field and 500 mG at the positive field. The FDP was asymmetrical at both fields, which is different from the expected symmetric FDP. This was the same as the asymmetric FDP for trapped flux measured at the vortex, as shown in Fig. 16d. It is deduced that the asymmetric *d*-wave FDP (Fig. 16c) is attributed to a trapped flux, instead of to the Josephson current with the $\pi$ phase difference.

Moreover, as for the bias current measured at the edge and corner SQUIDs (Figs. 16e and f), a ratio of the critical currents at *d*- and *s*-waves is given as $I_d/I_s \equiv \Delta R_s(\approx 2.3)/\Delta R_d(\approx 1.5) \approx 1.53$ to the theoretical value of $I_d/I_s \approx 1$ from $I=I_o|\cos((\pi\Phi/\Phi_o)+\delta/2)|$ where $\delta$ is zero for *s*-wave and $\pi$ for *d*-wave (Fig. 16f). The ratio is too large and similar to the value $I_d/I_s \approx 1.88$ (Figs. 16b and c) obtained in the trapped flux. This indicates that the observed half flux was attributed to the trapped flux, and not to supercurrent. Furthermore, it is suggested that the *s*-wave component exists in the sample on the ground of a constant value at zero field (Fig. 16c).

### 4-3-2. Flux image at superconducting ring on the tricrystal substrate

Subsequently, the half flux was also reanalyzed, and was imaged with a SQUID microscope and measured at the YBCO superconducting ring on the tricrystal substrate [89,90], as shown in Figs. 17a and b. The images have been suggested as strong evidence of *d*-wave symmetry. It has been assumed that the half flux is a result of the supercurrent generated by the circulation of the *d*-wave superconducting carriers in the superconductor ring. The observed flux seems to be evidence of *d*-wave symmetry. However, a detailed observation revealed that the flux image measured at the ring, including the tricrystal point, was anisotropic (or asymmetrical), indicating that the supercurrent was not circulated. From this, it follows that the anisotropic flux is not caused by supercurrent, but rather by a superconductor with trapped flux. Furthermore, for Figs. 17c and d, flux images for films covering the tricrystal point are also regarded as the trapped flux induced by a vortex coming from the structure of the tricrystal point, which has been confirmed [95]. Note that the trapped half flux caused by the vortex, not evidence of *d*-wave symmetry, differs from the half flux arising from the persistent supercurrent.

Moreover, Figure 17e shows a ring-type flux image measured in an optimal-doped single-layer tetragonal $Tl_2Ba_2CuO_{6+\delta}$ superconducting ring without any junction on the substrate [96]. The flux image shape is symmetric in intensity (color tone) in image, and is generated by supercurrent. This flux image is evidence of *s*-wave symmetry. Figure 17f shows a flux image of large intensity at the tricrystal point for the blanket film, which is regarded as the trapped flux in a vortex induced by the tricrystal structure.

Finally, the observed half fluxes showing an asymmetric behavior (effect) at the corner SQUID and at the superconducting ring with three junctions, including the tricrystal point, arise from trapped flux in the vortex and not from *d*-wave supercurrent. Therefore, it seems that the flux caused by supercurrent is $\Phi_o$, as shown in 17e. Thus, the cuprate superconductors have *s*-wave pairing symmetry.

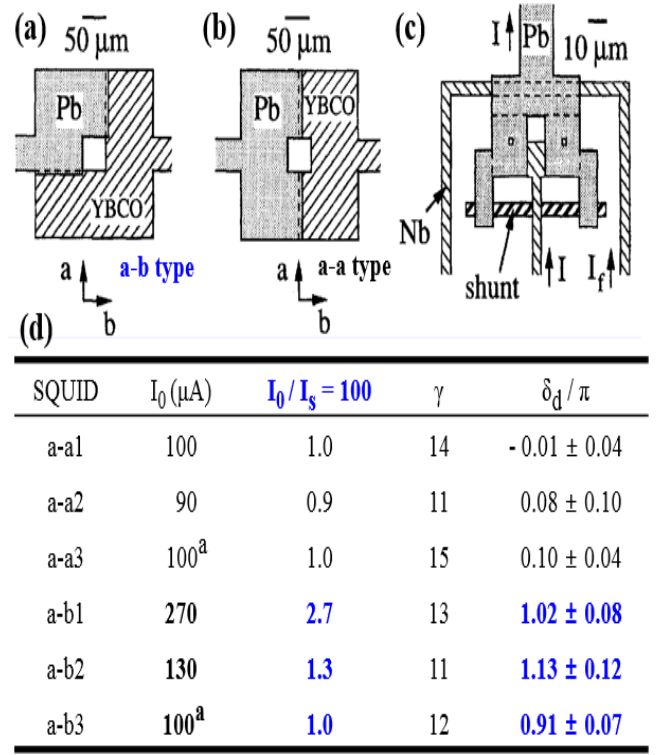

| SQUID | $I_0$ (μA) | $I_0/I_S = 100$ | $\gamma$ | $\delta_d/\pi$ |
|---|---|---|---|---|
| a-a1 | 100 | 1.0 | 14 | - 0.01 ± 0.04 |
| a-a2 | 90 | 0.9 | 11 | 0.08 ± 0.10 |
| a-a3 | 100[a] | 1.0 | 15 | 0.10 ± 0.04 |
| a-b1 | 270 | 2.7 | 13 | 1.02 ± 0.08 |
| a-b2 | 130 | 1.3 | 11 | 1.13 ± 0.12 |
| a-b3 | 100[a] | 1.0 | 12 | 0.91 ± 0.07 |

[a] Measured without magnetic shielding.

**Figure 18. Underdoped YBCO-Ag-Pb sample SQUIDs with the same dimensions.** (a) an a-b corner type, (b) an a-a corner type. (c) Nb-PbIn sensor SQUID. (d) Current $I_o$, ratio $I_o/I_s$, phase difference $\delta_d/\pi$ measured at a-a and a-b type SQUIDs for 6 samples. This was cited from Mathei *et al.* [91,92].

### 4-3-3. Current in SQUID and flux image of sample SQUID

For the underdoped YBCO-Ag-Pb sample SQUIDs with the same dimensions (Fig. 18a and b), using the Nb-PbIn sensor SQUID (Fig. 18c) [91,92], the phase differences of $\delta_d/\pi \approx 0.91 \sim 1.13$, proving the presence of half flux (evidence of $d$-wave), were observed for three samples with an a-b corner type, while, three samples were also measured with an a-a corner type, $\delta_d/\pi \approx 0.01 \sim 0.1$, regarded as the absence of the half flux (evidence of $s$-wave). The measured circulating currents, $I_o$, for the former were $I_{o\ d\text{-wave}} \approx 270$, 130, and 100 μA, and were $I_{o\ s\text{-wave}} \approx 100$, 90, and 100 μA for the latter. A ratio of $I_{o\ d\text{-wave}}/I_{o\ s\text{-wave}} \approx 2.7$, 1.3, and 1 was determined with a reference of $I_{o\ s\text{-wave}} \approx 100$, as shown in Fig. 18d (Table). The ratios was too large with respect to the standard value of 0.7 in section 4-3-1, indicating that the π shift arises from trapped flux.

In addition, Mathai, *et al.* [91,92] measured magnetic images of the sample SQUIDs using a scanning SQUID microscope with the sensor SQUID (Fig. 18c). For a SQUID of a-b type (Fig. 18a), the authors averaged positive and negative images to remove a trapped flux. The averaged image is shown in Fig. 19a. They suggested that the image shows a flux generated by a circulating supercurrent (Fig. 19b), as opposed to trapped flux, and that the measured flux at 40 nT is $-0.6\Phi_o$ corresponding to the half flux of n=-1/2. The grounds for the suggestion are that the image measured at the sample SQUID (Fig. 19a) is similar to the image measured at a ring-type superconductor (Fig. 19b) [92], because Mathai [92] assumed that the uniform closed black ring (Fig. 19b) was caused by supercurrent. However, in order to support the image (Figs. 19a and b), we calculated the magnetic flux induced by a constant current (presence of a supercurrent) with the Biot-Savart formula and noticed that the intensity of flux in circulating current is very large near the edge of the supercurrent ring, and that it monotonically decreases from the edge to center, where it is constant, as shown in Fig. 19c. Thus, because of the absence of a supercurrent trace (a change of contrast in black tone) near the ring edge in the images (Figs. 19a and b), we conclude that the measured closed-circle image shows a flux not the flux induced by the supercurrent, such as that measured in a vortex, as shown in Figs. 17c, d, and f. Generally, since the measured flux when a superconductor was trapped in a flux is asymmetry (or anisotropy), as explained in section 4-3-1, although the images are averaged, the trapped flux does not disappear. Furthermore, because Nb, Pb, and YBCO superconductors in the sample and sensor SQUIDs easily trapped the magnetic flux [97], it would be difficult to believe that the measured half flux arises from $d$-wave symmetry supercurrent.

### 4-3-4. Flux trap in SQUIDs and paramagnetic Meissner effect

Angular dependence of order-parameter symmetry in underdoped $YBa_2Cu_3O_{7-\delta}$ was measured at 4.2 K using a scanning SQUID microscope composed of twinned thin-film $YBa_2Cu_3O_{7-\delta}$ (YBCO)-Ag-PbIn SQUIDs [93,94]. The measurements were performed at 13 different tunneling angles and the observed order parameter in YBCO had time-reversal symmetry, and showed both $d_{x^2-y^2}$-pairing symmetry with π-phase shift and $s$-wave symmetry without phase shift, the latter being less than 5%. Any time-reversal symmetry breaking less than 5% can be ruled out, such as $d_{x^2-y^2} + is$ or $d_{x^2-y^2} + id_{xy}$. Although $s$-wave symmetry was observed, the authors suggested that the order parameter in YBCO is $d_{x^2-y^2}$ pairing symmetry, because, if YBCO had $s$-wave symmetry, the π-shift would not be expected, regardless of the dependence of the tuning angle or the usefulness of the junction preparation technique. When the sample SQUID was made, the YBCO films were chemically etched and milled with Ar ions. This fabrication process damages the films showing the intrinsic characteristics, easily forming the pseudogap phase as an extrinsic semiconductor with impurities in YBCO. Therefore, it follows from the analysis results that the π-shift $d$-wave symmetry comes from the pseudogap phase with the pseudogap at an antinode.

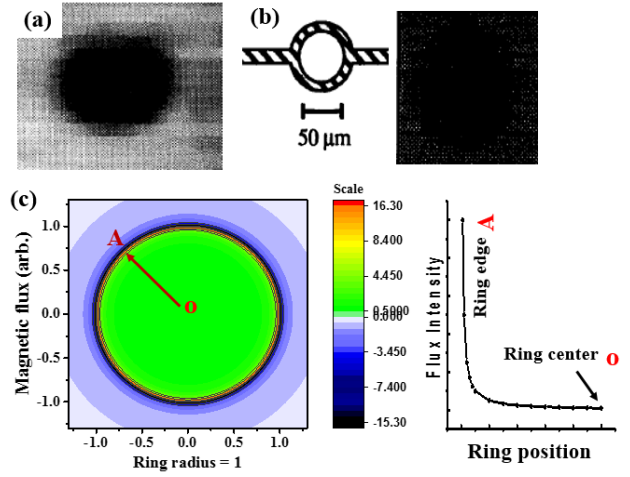

**Figure 19. Trapped flux in a superconducting ring.** (a) An average of positive and negative images measured for a SQUID of a-b type (Fig. 18a) [91,92]. The average was performed to remove the trapped flux. (b) Image measured at a ring-type superconductor cited from Mathai's PhD dissertation paper [92]. (c) is the simulation image and data induced by a circulating current. The simulation used the Biot-Savart formula by assuming that a $d$-wave supercurrent exists. The intensity of flux is very large near the edge of the supercurrent ring, and monotonically decreases from the edge to the center, and is constant at the center of ring. Outside of ring is ignored because outside flux when there is a persistent supercurrent does not exist.

Moreover, using a 4.2 K scanning SQUID microscope, twined thin-film underdoped $YBCO_{7-\delta}$-Ag-PbIn SQUIDs were examined and the phases of the order parameters for 13 different tunneling angles were measured [93,94]. The measured π-shift was obtained for 40°, 60°, 100°, 120°, 130°, 135°, 150°, and 160°, and the 0-shift caused by Meissner effect was obtained for 30°, -10°, -45°, -50°, and -140°. Based on the experimental results, the author did not rule out a real superposition of $s$ and $d_{x^2-y^2}$ [93,94] although the study did report an observation of $d$-wave symmetry. Further, the author attempted to remove trapped flux for some sample SQUIDs by means of increasing the sample temperature above $T_c$. However, the trapped flux was not perfectly removed [94] and it was noted that the method used to increase the sample SQUID temperature above $T_c$ may not have been capable of perfectly removing it. Generally, the scanning magnetic microscope had the YBCO sample

SQUID, the sensor SQUID with Nb and Pb superconductors, and the Nb superconducting coil applying the magnetic field. The superconductors with Nb and Pb and an underdoped $YBCO_{7-\delta}$ easily trapped the magnetic flux [97]. Therefore, in order to measure the pairing symmetry of the order parameter, all of the trapped fluxes in the sample and sensor SQUIDs, as well as the superconducting field coil, should be perfectly deleted, but this is no easy task. We conducted experiments to remove the trapped flux for BKBO superconductors [98]. The trapped flux could be not removed without the evaporation of all of the liquid He at room temperature. After putting new liquid He in the dewar of the system, the diamagnetic susceptibility in the virgin field applied into the sample and sensor SQUIDs was measurable. Nevertheless, we found that the magnitude of the diamagnetic susceptibility measured by zero field cool disagreed with that observed by field cool. Thus, regardless of the removal method, it is concluded that the flux is trapped and cannot be perfectly removed due to inhomogeneity of superconductor components.

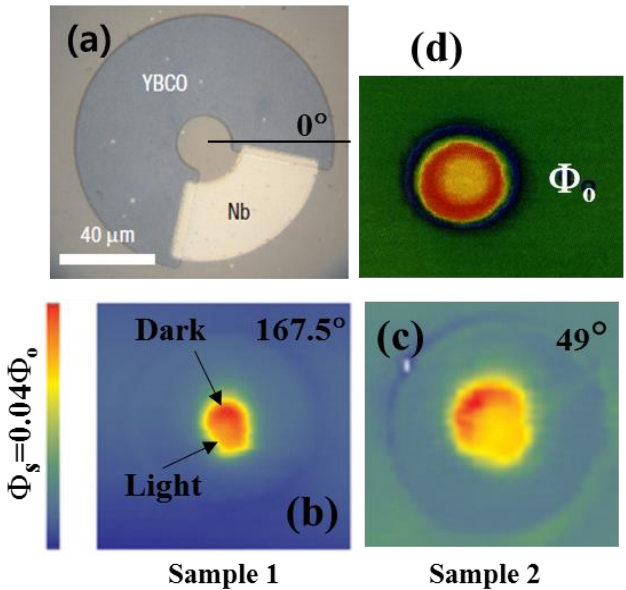

**Figure 20. Flux images measured at samples 1 and 2.** (a) Optical micrograph image of one of rings of a YBCO-Nb Ramp-edge junction in experimental samples 1 and 2. (b) A flux image taken at 167.5° with a high-spatial-resolution scanning SQUID microscope at sample 1. Slight inhomogeneous color tones are shown. (c) A flux image pictured at 49° in YBCO sample 2. A strong asymmetry in color tone is shown. (d) A flux image well measured in a superconducting ring for $Tl_2Ba_2CuO_{6+d}$. The color tone is uniform. This flux image is placed for a comparison of the flux images of Figs. (b) and (c). Figs. (a-c) was cited in Kirtley *et al.* [99]. Fig. (d) was cited in Tsuei *et al.* [96].

Additionally, unlike the previous research performed in underdoped samples, it is introduced that the angle-resolved phase-sensitive experiments measured in optimally doped YBCO samples for determination of the in-plane gap symmetry [99]. The optimally doped samples have the highest $T_c$ and the well-developed Fermi surface of ring type in ARPES, and shows the s-wave behaviors of the U-shape gap in tunneling experiments. Two samples in the experiment were used. Each sample was composed of 72

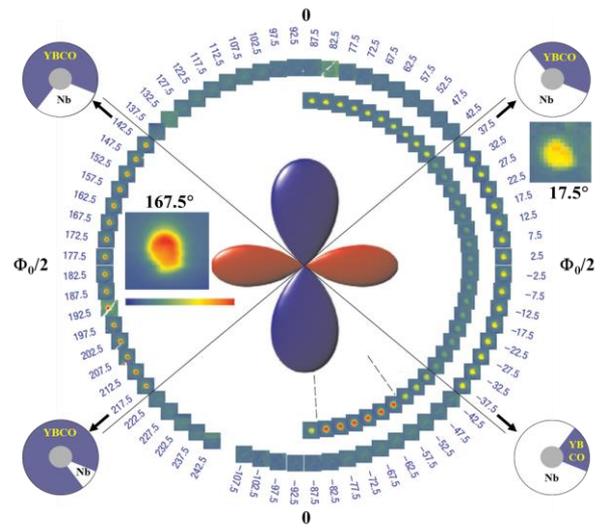

**Figure 21. Angle dependence of flux images** measured by the scanning SQUID microscope for YBCO sample 1. The images showing $\Phi_o/2$ at left side, measured by rings with a wide YBCO area, are more strong that those at right side. This was cited in Kirtley *et al.* [99].

YBCO semi-rings with the highly untwinned (85%) YBCO/Au/Nb (s-wave) ramp-edge Josephson junctions [99]; the YBCO ring image is shown in Fig. 20a. The flux images were measured by a high-spatial-resolution scanning SQUID microscope with a pickup loop made of $Nb/AlO_x/Nb$ trilayer junctions. The Josephson-π shift with the half flux($0.5\Phi_0$) proposed as evidence of *d*-wave symmetry as well as the zero shift were measured with angle. Authors [99] reported that the *d*-wave component to the *s*-wave one exists for approximately 90% on the basis of the fitting of the flux measured in sample 2. Moreover, although the image for sample 1 shows slightly anisotropy in color tone, this has been suggested as evidence of the π junction. Therefore, in order to reveal the presence of the half flux, it should be analyzed whether the half flux in images is generated by *d*-wave supercurrent or not. Firstly, more clearly, we compare the image of sample 1 (Fig. 20b) with the flux image which is well-measured and induced by supercurrent for $Tl_2Ba_2CuO_{6+\delta}$ [96] (Fig. 20d). The image of sample 1 is quite different from the image which is nearly uniform in the color tone in Fig. 20d. Secondly, as shown in Fig. 21, although it was measured at the same sample 1, the magnitude of the measured flux at the left and right sides is different. The intensity of flux at left side is stronger than that at right side. The left rings have a much larger area of YBCO than right rings. The intensity of flux seems to depend on the extent of the YBCO area. The dependence may be because inhomogeneity in YBCO is larger than that in Nb, which can be a cause for flux to be easily penetrated into the superconductor [98]. Therefore, the inhomogeneity and the anisotropic behavior allude that the measurement of the absolute value of the half flux is difficult. Note that the image (Fig. 20b) can have a ring flux of *s*-wave type except for the anisotropic behavior because the optimally doped sample has well-developed Fermi surface. As for sample 2, the flux image observed (Fig.20c) shows an obvious anisotropy which is the characteristic of a trapped flux to a superconductor (Fig.16d). Thus, these conclude that the half flux in the images is not a flux induced by the circulation of *d*-wave spontaneous supercurrent due to an anisotropic effect in intensity (color tone). If the anisotropic effect is removed in the images, the *s*-wave flux would remain, for instance such as the image (Fig. 20d). Moreover, the presence of a *d*-wave component by 90% mentioned previously is not proved. It is also deduced that, although the optimally doped samples were used, the Fermi surface of ring type was not developed.

Furthermore, the paramagnetic-Meissner effect (PME) was proposed as another piece of evidence for the Josephson-π junction with the half-flux quantum [100], and was suggested in support of the idea that cuprate superconductors have an order parameter of *d*-wave symmetry. The PME was observed in BKBO [98], BSCCO [101], YBCO [102], $MgB_2$ [103], and Nb [104,105]. Generally, because a SQUID measuring the PME tends to trap magnetic flux, a trapped SQUID sensor always measures the PME, irrespective of *d*-wave or *s*-wave type superconductors [98], although a virgin SQUID sensor measures the diamagnetic Meissner effect in a superconductor [98]. It has also been reported that the PME stems from such extrinsic effects as the flux trap [97,98,101,105,106].

**4-3-5. Summary of experiments**

From the research presented here, it follows that the semiconductive pseudogap in (underdoped or optimally) superconductors with an incompleted Fermi surface exists at the antinode and the superconducting gap exists at the node. The magnetic flux (or field) is penetrated from the antinode and expelled (or shielded) at the node. The penetrated flux is trapped by the superconductor at the node. Therefore, the magnetic flux is necessarily trapped, regardless of attempts to remove it. Therefore, the anisotropic or asymmetric effect in the measured half-flux-quantum image or data shown in previous papers supports the fact that the magnetic flux in the sample was trapped, even though the magnetic field is small. We assert that the anisotropic (or asymmetric) effect is caused by the incompleted Fermi surface. Accordingly, for underdoped or optimally samples with the incompleted Fermi surface, it is concluded that the pairing symmetry (or order parameter) is not accurately revealed simply by measuring magnetic flux such as the Josephson junction, the SQUID, or the Paramagnetic Meissner effect, using the magnetic field.

**Conclusion**

The analysis of the measured diverging effective mass fitted by the Kim-DEM (Fig. 2a) revealed that, as the hole doping concentration increased, the Fermi surface grew from the nodal Fermi point to the anti-nodal point, which forms the isotropic Fermi surface with *s*-wave symmetry-type characteristics and hole doping. This finding indicates that the underdoped state is intermediate to the isotropic state. The high-$T_c$ mechanism is caused by a strong correlation ($\kappa_{BR}$>0.9) induced by the maximum carrier density at the nodal Fermi point ($\rho\approx1$), generated by the *d*-wave MIT at the node in the pseudogap $CuO_2$ layer. The superconducting intrinsic gap is formed at the quantum critical point of the node and grows from the node (underdoped) to the anti-node (optimal doped), which ultimately shows *s*-wave symmetry.

Accordingly, the analysis of the diverging effective mass revealed important unsolved problems in cuprate superconductors, such as the nodal constant Fermi energy [25,28], the relationship between the pseudogap and the superconducting gap, the identity of pairing symmetry, and the reason that cuprate layer superconductors have higher-$T_c$ than low-$T_c$ 3-dimensional superconductors. Our analyses revealed the absence of experimental results to show $d_{x^2-y^2}$-wave Cooper-pair pairing symmetry. Furthermore, superconductivity in hydrogen sulfide was recently discovered under high pressure below the new record of high $T_c$ of 203 K [107], and this material, presumably crystalline $H_3S$, has a conventional isotope effect as demonstrated by doping the H with D. Thus, it is clearly a conventional BCS superconductor, so the many arguments that a superconductor with a $T_c$ in excess of 40 K had to have an "unconventional" pairing mechanism have been definitively proven incorrect. Furthermore, the high-$T_c$ superconductor can be explained in the context of the BCS theory.

Note that parts of this research were presented at APS March meetings from 2014 to 2017, $M^2S$ HTSC 2015, JPS Meetings in 2016 and 2017, and an international conference (ECSN-2017) held at Odessa in 2017.

**Competing Interests**

The author(s) declare(s) that there is no conflict of interest regarding the publication of this paper.

**Acknowledgements**

We introduced this research to Prof. S. Sebastian (an author in Ref. 32) at 2014 APS March meeting (Denver) and received correct comments on experiments. We acknowledge Prof. T. Tohyama, Prof. D. N. Basov and Prof. M. M. Qazilbash for valuable comments and Prof. T. Yoshida, Dr. M. Hashimoto, Prof. H. Yagi, Dr. S. Yasmine, Prof. S. Sebastian for reading, comments and permission to use data. We also acknowledge Prof. D. J. Van Harlingen, Dr. C. C. Tsuei, Prof. F. C. Wellstood, Prof. M. H. Hamidian, Prof. J. Lee., and Prof. J. C. Seamus Davis for using their data. Figure 19c was provided by Sung-Woo. Cho calculated the Biot-Savart formula. Drawings of Figures 6(c), 9 and 14 were helped by Sun-Kyu Jung and Tetiana Slusar. This research was supported by both the creative project (17ZB1220) in ETRI and a MS&ICT project (2017-0-00830) on MIT. We received permissions for using of Figs.11(a-f), Fig. 12b, Fig. 13, Fig. 15, Fig. 16, Fig. 17, Fig. 18, Figs. 19(a and b), Fig. 20, Fig. 21 from journals presented the figures.